\documentclass[preprint,12pt]{elsarticle}

\usepackage{float}

\usepackage{amssymb}
\usepackage{amsmath}
\usepackage{comment}
\usepackage{caption}
\usepackage{booktabs} 
\usepackage{siunitx}  
\usepackage{color}
\usepackage[dvipsnames,svgnames]{xcolor}
\usepackage{hyperref}
\usepackage{cleveref}
\usepackage{multirow}

\usepackage{listings}
\lstdefinestyle{bashstyle}{
  language=bash,
  basicstyle=\ttfamily\small,
  numbers=left,
  numberstyle=\tiny,
  stepnumber=1,
  numbersep=8pt,
  frame=single,
  breaklines=true,
  breakatwhitespace=false,
  showstringspaces=false,
  tabsize=2
}

\newcommand{\bfxi}[0]{\boldsymbol{\xi}}

\newcommand{\bfd}[0]{{\boldsymbol{d}}}

\newcommand{\bfe}[0]{{\boldsymbol{e}}}
\newcommand{\bfF}[0]{{\boldsymbol{F}}}

\newcommand{\bfI}[0]{{\boldsymbol{I}}}

\newcommand{\bfn}[0]{{\boldsymbol{n}}}

\newcommand{\bfp}[0]{{\boldsymbol{p}}}

\newcommand{\bfu}[0]{{\boldsymbol{u}}}

\newcommand{\bfx}[0]{{\boldsymbol{x}}}
\newcommand{\bfX}[0]{{\boldsymbol{X}}}

\newcommand{\bfsigma}[0]{\boldsymbol{\sigma}}

\newcommand{\bfvarepsilon}[0]{\boldsymbol{\varepsilon}}

\newcommand{\bflambda}[0]{{\boldsymbol{\lambda}}}

\newcommand{\hatn}[0]{\hat{\bfn}}
\newcommand{\hatbfu}[0]{\hat{\bfu}}
\newcommand{\hatbfx}[0]{\hat{\bfx}}
\newcommand{\hatbflambda}[0]{\hat{\bflambda}}
\newcommand{\khat}[0]{\widehat{K}}

\newcommand{\ical}[0]{\mathcal{I}}
\newcommand{\jcal}[0]{\mathcal{J}}

\newcommand{\lcal}[0]{\mathcal{L}}

\newcommand{\pcal}[0]{\mathcal{P}}
\newcommand{\qcal}[0]{\mathcal{Q}}

\newcommand{\tcal}[0]{\mathcal{T}}
\newcommand{\ucal}[0]{\mathcal{U}}
\newcommand{\vcal}[0]{\mathcal{V}}
\newcommand{\xcal}[0]{\mathcal{X}}

\newcommand{\utesti}[0]{\boldsymbol{\varphi}_{\bfu,i}}
\newcommand{\ptesti}[0]{\varphi_{p,i}}
\newcommand{\ltesti}[0]{\boldsymbol{\varphi}_{\bflambda, i}}
\newcommand{\xtesti}[0]{\boldsymbol{\varphi}_{\bfx,i}}

\newcommand{\ltestj}[0]{\boldsymbol{\varphi}_{\bflambda,j}}

\newcommand{\Omegaz}[0]{{\Omega_0}}
\newcommand{\Omegat}[0]{{\Omega_t}}
\newcommand{\Gammat}[0]{{\Gamma_t}}
\newcommand{\intom}[1]{\int_\Omega{#1}\,d\bfx}
\newcommand{\intomt}[1]{\int_{\Omega_t}{#1}\,d\bfx}

\newcommand{\intdomt}[1]{\int_{\partial\Omega_t/\Gamma_t}{#1}\,d\bfx}
\newcommand{\intomzero}[1]{\int_{\Omega_0}{#1}\,d\bfX}
\newcommand{\intdomzero}[1]{\int_{\partial\Omega_0}{#1}\,d\bfX}
\newcommand{\intgam}[1]{\int_\Gamma{#1}\,d\bfx}
\newcommand{\intgamt}[1]{\int_{\Gamma_t}{#1}\,d\bfx}

\newcommand{\NL}[0]{\text{NL}}

\newcommand{\dealii}[0]{{\texttt{deal.II}} }

\newcommand{\R}[0]{{\mathbb{R}}}

\newcommand{\Rd}[0]{{\mathbb{R}^d}}

\DeclareMathOperator{\trace}{tr}

\crefformat{equation}{(#2#1#3)}
\crefrangeformat{equation}{(#3#1#4) to~(#5#2#6)}
\crefmultiformat{equation}{(#2#1#3)}%
{ and~(#2#1#3)}{, (#2#1#3)}{ and~(#2#1#3)}

\begin{document}

\begin{frontmatter}

\title{A parallel solver for vortex-induced vibrations at zero mass ratio}

\author[1]{A. Bawin\corref{cor1}}
\ead{arthur.bawin@polymtl.ca}
\author[1]{J. Le Pouhaër}
\ead{joan.le-pouhaer@etud.polymtl.ca}
\author[1]{S. Étienne}
\ead{stephane.etienne@polymtl.ca}
\author[1]{C. Béguin}
\ead{c.beguin@polymtl.ca}
\cortext[cor1]{Corresponding author}

\affiliation[1]{organization={Département de génie mécanique,
Polytechnique Montréal},
addressline={2500 Chem. de Polytechnique},
city={Montreal},
citysep={}, 
%
postcode={H3T 0A3},
state={QC},
country={Canada}}

\begin{abstract}
We present a parallel strategy for the study of the vortex-induced vibrations (VIV) of
a rigid, massless cylinder mounted on translational springs, in two and three dimensions.
Without mass or damping, its motion is governed solely by the fluid forces
and the spring restoring force, and is an immediate response to the flow.
We focus on low spring stiffness,
yielding high reduced velocity and maximal oscillation amplitude.
A monolithic finite element method in an arbitrary Lagrangian-Eulerian framework
with implicit time stepping is proposed,
solving the fluid flow, cylinder dynamics, and mesh motion simultaneously.
Mesh movement is modelled with a linear elastic pseudo-solid with
variable Lamé coefficients, and 
the no-slip condition on the cylinder is enforced with a Lagrange multiplier
coupled to the position degrees of freedom;
three strategies are explored to enforce this
force-position coupling in a distributed framework.
To the best of our knowledge, this work is the first to propose a distributed solver
for three-dimensional fluid-structure interaction at zero mass ratio.
The solver is verified with manufactured solutions,
and exhibits second-order accuracy in time.
A numerical study of two- and three-dimensional VIV for Reynolds numbers from 100 to 300
reproduces the VIV responses obtained in previous two-dimensional
studies, and predicts nonperiodic 3D oscillations starting at $Re = 300$.
Using the direct solver MUMPS, the proposed method scales to up to 768 compute cores.
\end{abstract}


\end{frontmatter}

\section{Introduction}

\paragraph{Motivation.}
Vortex-induced vibrations (VIV) are extensively studied fluid-structure interaction
(FSI) phenomena,
arising from the interaction between the unsteady vortex shedding
in the wake of a bluff body and the body's structural motion
\cite{williamson2008brief, sarpkayaWaveForcesOffshore2010, paidoussisFluidStructureInteractions2010, huera2025vortex}.
These vibrations lead to potentially large oscillations,
and can cause a synchronization of the structure with the vortex shedding frequency,
called 'lock-in' regime.
In marine engineering, to give only one example, vortex-induced
vibrations
cause fatigue loads and affect the lifespan of offshore structures
such as buoys, risers, and mooring systems. Accurately predicting the maximal amplitude
of these oscillations is thus desirable for the design and safety of these structures.
For a rigid structure, Yu et al. \cite{yu2018two}
proposed numerical simulations of two-dimensional VIV at
low Reynolds number ($Re = 175$), and showed that the peak amplitude is obtained at
zero solid-to-fluid mass ratio $m^* = \rho_\text{solid}/\rho_\text{fluid}$, motivating the numerical exploration of VIV in this regime.

From a numerical standpoint, the simulation of VIV is challenging due to the
strong coupling between the fluid and structural dynamics
\cite{causinAddedmassEffectDesign2005a},
and even more so when considering bodies with low mass ratios \cite{jaimanStableSecondorderPartitioned2016}.
In the limit of a system with zero mass ratio, the structural motion
is entirely governed by the fluid forces;
the solid body has no inertia, instead, the effective inertia of the coupled system
is due to the added-mass mechanism \cite{causinAddedmassEffectDesign2005a, banksStablePartitionedFSI2017a}.
Traditional
partitioned algorithms
split the FSI problem into two distinct fluid and structure subproblems.
\emph{Strongly-coupled} partitioned methods successively solve for the fluid domain,
postprocess the fluid forces on the solid body,
then update the body and mesh position
in a fixed point loop, until convergence of both the flow
and the body motion.
\emph{Weakly-coupled} methods, on the other hand, do not perform this retroaction loop \cite{ha2020investigation}.
Partitioned methods are typically easier to implement and parallelize,
and allow to use existing dedicated fluid and solid solvers.
They require, however, large enough mass ratios and small enough time steps to
ensure robustness,
and typically become unstable for light bodies.
In the zero mass limit,
the added-mass contribution must be incorporated in the interface update
to maintain a stable numerical scheme
\cite{michler2004monolithic, causinAddedmassEffectDesign2005a,banksStablePartitionedFSI2017a,banksStablePartitionedFSI2017b}.
Contrary to partitioned methods, monolithic formulations
solve the fluid, interface and solid body equations in a single,
fully coupled system, yielding robust schemes with consistent time accuracy, even at zero mass ratio
\cite{etienneLowReynoldsNumber2012, couture2020new}.
They are commonly formulated in the Arbitrary
Lagrangian-Eulerian (ALE) framework \cite{donea2004arbitrary}, 
in which the fluid, structure, and mesh dynamics
are coupled within a single nonlinear system,
see for instance \cite{hron2006monolithic, wick2013solving, langer2016recent, failer2020parallel, wangEnergyStableOnefield2020, sun2025parallel}
for 2D and 3D applications,
and \cite{etienne2009perspective, etienneLowReynoldsNumber2012, couture2020new}
for 2D applications at zero mass ratio.
In the particular case of a rigid solid body,
\cite{yu2018two} also proposed an approach where the incompressible Navier-Stokes equations
are formulated in an ALE point of view and in a moving frame, suitable for the simulations of VIV at zero mass ratio.
This avoids computing the mesh position or displacement, but makes the interpretation of
the boundary conditions less straightforward.

\paragraph{VIV at zero mass ratio.}
To our best knowledge, the numerical investigation of vortex-induced vibrations at zero mass ratio
has been for now limited to two-dimensional cases,
based on the works of \cite{shiels2001flow, etienneLowReynoldsNumber2012, yu2018two, couture2020new}.
Using a viscous vortex boundary method, Shiels et al. \cite{shiels2001flow} studied the transverse oscillations (one degree of freedom) of a cylinder at $Re = 100$ without mass and damping, yielding peak amplitudes of $0.59D$.
Étienne et al. \cite{etienneLowReynoldsNumber2012} and Yu et al. \cite{yu2018two}
considered a cylinder mounted on translational springs of stiffness~$k$ (two degrees of freedom),
free to oscillate in the $xy$-plane,
and reported peak amplitudes of $0.9D$.
In these works, the hydrodynamic forces are recovered
from the reactions associated with strongly imposed Dirichlet boundary conditions,
requiring ad-hoc matrix manipulations.
Couture et al. \cite{couture2020new} investigated the rotation of a massless
cylinder with elliptic cross-section, and
used a Lagrange multiplier to weakly enforce the no-slip condition on the cylinder.
The Lagrange multiplier is identified to the force density on the cylinder,
allowing to recover the efforts implicitly and without matrix manipulation,
at the cost of an additional variable on the boundary of the cylinder.
Numerical simulations of three-dimensional VIV at zero mass ratio remain
largely unexplored, due, on the one hand, to the increase of computational resources
needed to solve large nonlinear monolithic systems,
and, on the other hand,
to the lack of solvers to simulate these phenomena.

\paragraph{Overview of the proposed methodology.}
The objective of the present work is to 
address the lack of solvers for the simulation of FSI at zero mass ratio,
and extend these previous developments toward
robust, large-scale, distributed simulations of VIV around
massless solid bodies
in three dimensions.
To this end, we propose an
implicit and monolithic finite element formulation in an ALE framework,
involving a Lagrange multiplier approach
for the implicit evaluation of hydrodynamic forces, and a
pseudo-solid mesh movement model.
Distributed memory parallelism is handled by both the \dealii library \cite{arndt2025deal},
and on our end, by handling the couplings between the Lagrange multiplier and mesh position degrees of freedom
consistently across mesh partitions.
A particular attention is devoted to the treatment of the rigid-body force-position
coupling induced by the motion of the cylinder, for which three distributed coupling strategies are compared.
The proposed methodology is verified
with
manufactured solutions,
and the flow solver with a Lagrange multiplier is validated against benchmark flows
around bluff bodies.
As the zero mass ratio is an idealized model, the complete FSI solver cannot be validated against experimental data, however.
Numerical methods for the parallel resolution of the linear systems arising from the discretization of
monolithic and nonlinear FSI systems (e.g., Newton-Krylov methods) have been discussed
in, e.g., \cite{langer2016recent, failer2020parallel, sun2025parallel}.
To prototype the implementation of the fluid-structure coupling at zero mass ratio,
however,
our solver currently relies on the parallel sparse direct solver MUMPS \cite{amestoy2000mumps},
which is used as silver bullet, but limits the scaling of the proposed solution. The implementation of efficient preconditioners
will be considered in an upcoming work.

We consider, as in \cite{etienneLowReynoldsNumber2012, yu2018two},
the uniform flow with velocity $U_\infty$ around a cylinder with diameter $D$ mounted on translational springs in the $xy$-plane,
and blocked along the $z$-axis.
In addition to the Reynolds number $Re = U_\infty D/\nu$ and the mass ratio $m^*$,
VIV in these conditions are characterized by the reduced velocity
$U_r := U_\infty / (f_n D)$, the ratio of the fluid and structure characteristic velocities,
with $f_n = \sqrt{k/m_a}/(2\pi)$ the natural frequency of the massless cylinder.
In the expression of $f_n$, $m_a = \rho\pi L D^2/4$ denotes the added mass induced
by a surrounding fluid of density $\rho$, with $L$ the cylinder length \cite{etienneLowReynoldsNumber2012}.
Thus, the reduced velocity for a massless solid body writes $U_r = \pi U_\infty \sqrt{\pi \rho L/k}$ \cite{yu2018two}.
We are interested in the parameter space described by Reynolds numbers ranging from 100 to 300,
and reduced velocities between 0 and 12,
allowing to extend previous two-dimensional studies toward flow regimes exhibiting
significant three-dimensional wake dynamics.

\label{sec:introduction}

\paragraph{Outline.}
This paper is organized as follows.
The geometry, governing equations, and boundary conditions of the FSI problem are introduced in \Cref{sec:equations},
and \Cref{sec:numerical_scheme} presents the adopted finite element weak formulation and numerical scheme.
Because of the moving grid, additional terms must be accounted for in the Jacobian
matrix arising from the Newton-Raphson method: this is briefly discussed in \Cref{sec:jacobian}.
In \Cref{sec:implementation}, we discuss possible coupling strategies between the position and Lagrange multiplier
degrees of freedom on the cylinder, which are critical to obtain acceptable solve times.
In \Cref{sec:v_and_v}, the overall method is verified with manufactured solutions,
and the flow solver is validated against experimental data of laminar flows around a cylinder and a sphere.
The paper concludes on a numerical study of vortex-induced vibrations around a
massless cylinder in two and three dimensions in \Cref{sec:results}.
Our two-dimensional results are compared to those of Yu et al. \cite{yu2018two}
for verification, then we investigate the velocity and vorticity patterns of three-dimensional VIV
for Reynolds numbers between 100 and 300.
We conclude with scaling considerations, which are currently limited by the use of
a direct solver, and propose ideas for further improvements.

\section{Governing equations}
\label{sec:equations}
\begin{figure}
  \centering
  \includegraphics[width=\linewidth]{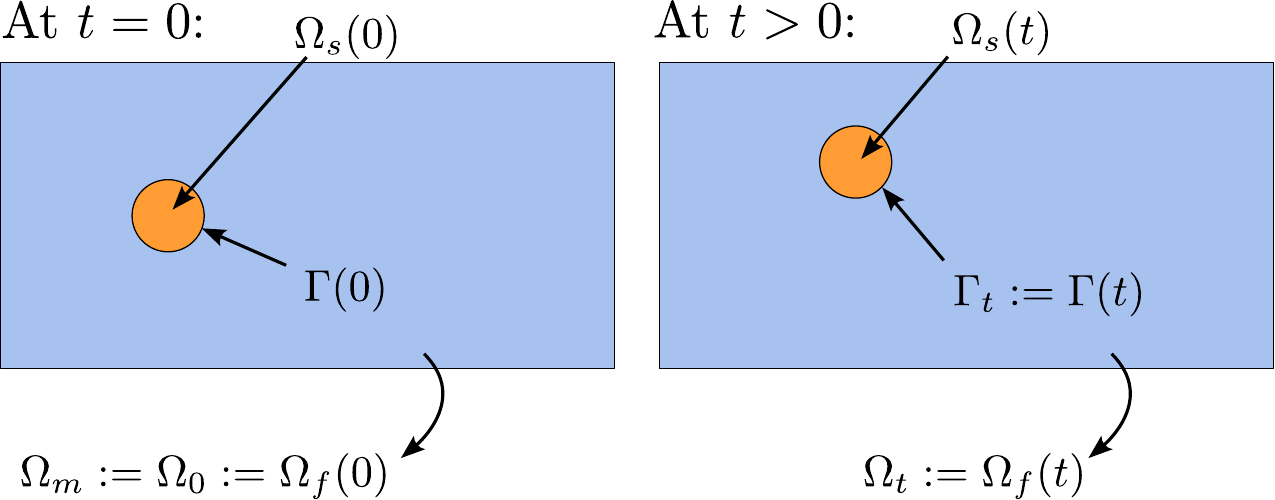}
  \caption{Geometry of the fluid-structure interaction system.}
  \label{fig:geometry}
\end{figure}

We consider a bounded region $\Omega \subset \Rd$ with $d = 2, 3$,
consisting of non-overlapping, time-dependent fluid and solid domains, noted $\Omega_{f}(t)$ and $\Omega_s(t)$ respectively.
The equations of interest are the incompressible Navier-Stokes equations,
describing the dynamics of the flow,
and the quasi-static momentum balance of the solid region.
To account for the movement of the solid, we consider a moving mesh method
and an arbitrary Lagrangian-Eulerian (ALE) formulation of the Navier-Stokes equations
in an inertial frame of reference.
The mesh movement is modeled with a linear elastic analogy
solved in a purely Lagrangian formulation,
which forms the third equation of the system,
solved on the fixed domain $\Omega_m := \Omega_f(0)$.
We denote by $\bfX$ the coordinates in the reference configurations (i.e., at $t = 0$),
on which the solid and mesh movement equations are solved,
and by $\bfx = \bfx(\bfX, t)$ the coordinates in the current configurations.
The position field $\bfx(\bfX, t)$ is solution of the mesh movement equation,
and defines the fluid domain $\Omega_{f}(t)$ at all times, \Cref{fig:geometry}.
The boundary of the solid domain is noted $\Gamma_t := \Gamma(t)$.

Because we consider a rigid cylinder, the momentum balance in the solid region
reduces to a balance of external forces at its center $\bfX_c$.
For a massless cylinder mounted on a spring with constant and isotropic stiffness
$k$, this balance involves only the fluid forces and the restoring force of the spring:
\begin{equation}
  \label{eq:solid_balance}
  \bfF_{\text{ext} \to \text{solid}} = \bfF_{\text{spring}} + \bfF_{\text{fluid}} = -k(\bfx_c(\bfX_c, t) - \bfX_c) + \bfF_{\text{fluid}} = 0,
\end{equation}
Thus, the solid equation degenerates into a constraint on the position of its center
of mass, which is treated hereafter as a boundary condition for the mesh movement equation.
As a result, only the fluid and mesh equations are to be actually solved,
in an ALE and Lagrangian point of view respectively,
and in the following we simply denote their respective domain by $\Omega_t := \Omega_f(t)$
and $\Omega_0 := \Omega_m = \Omega_f(0)$ without ambiguity.

Denoting by $\bfu(\bfx,t)$ and $\bfp(\bfx,t)$ the flow velocity and pressure respectively,
the considered system of equations writes:
\begin{equation}
\begin{alignedat}{2}
\rho \left( \partial_t{\bfu} +
\left(\left(\bfu - \partial_t{\bfx}\right)\cdot \nabla\right)\bfu \right)
- \nabla \cdot \bfsigma_f(p, \bfu) &= 0 &&~~~~\text{in } \Omega_t,\\
-\nabla \cdot \bfu &= 0 &&~~~~\text{in } \Omega_t,\\
-\nabla_{\bfX} \cdot \bfsigma_s(\bfx) &= 0 &&~~~~\text{in } \Omega_0.
\label{eq:system}
\end{alignedat}
\end{equation}
where $\nabla := \nabla_{\bfx} = \partial_{x_i}$ denotes the gradient in
the current configuration, $\nabla_{\bfX} := \partial_{X_i}$ the gradient in the
reference configuration, and with the linear stress tensors defined by:
\begin{equation}
  \begin{alignedat}{3}
    \bfsigma_f(p, \bfu) &= ~~~~\,-p \bfI &&+ 2\mu_f \,\bfd(\bfu),
    ~~~~\bfd(\bfu) &&= (\nabla \bfu + \nabla \bfu^T)/2,\\
    \bfsigma_s(\bfx) &= \lambda_s \trace(\bfvarepsilon)\bfI &&+ 2\mu_s \,\bfvarepsilon(\bfx),
    ~~~~~\bfvarepsilon(\bfx) &&= (\nabla_{\bfX} \bfx + \nabla_{\bfX} \bfx^T)/2 - \bfI.
  \end{alignedat}
\end{equation}
The first and second equations of \cref{eq:system} are the ALE form of the
incompressible Navier-Stokes equations for a fluid with constant density $\rho$
and dynamic viscosity $\mu_f$,
whereas the third equation is the quasi-static momentum balance for the pseudo-solid
used to define the mesh movement, with Lamé coefficients $\lambda_s, \mu_s$.
In this steady-state equation, the position field is the instantaneous mesh response
to the time-varying boundary condition obtained from the force balance \eqref{eq:solid_balance}.
In our formulation, the pseudo-solid equation is solved for the mesh \emph{position} instead of the mesh
displacement $\bfx - \bfX$, as this leads to a more natural manipulation of position-dependent
finite element mappings, especially within the \dealii library, see \Cref{sec:fem}.

System \cref{eq:system} is completed with the
following initial and boundary conditions on the fields $\bfu(\bfx,t), p(\bfx,t)$ and $\bfx(\bfX, t)$.
At $t = 0$, the initial velocity and mesh position are prescribed:
\begin{equation}
\bfu(\bfx, 0) = \bfu_0(\bfx) \text{ and } \bfx(\bfX, 0) = \bfX \text{ in }\Omega_0,
\end{equation}
and at $t > 0$, the following Dirichlet and Neumann conditions are applied.
For the flow problem, we set:
\begin{equation}
  \label{eq:flowBC}
\begin{alignedat}{2}
  \bfu &= \bfu_{\text{in}} &&\text{ on }\partial\Omega_\text{inlet}(t),\\
  \bfu \cdot \hatn &= 0 &&\text{ on }\partial\Omega_\text{slip}(t),\\
  \bfu             &= \partial_t\bfx &&\text{ on }\partial\Omega_\text{no slip}(t) = \Gamma_t,\\
  \bfsigma(p, \bfu) \cdot \hatn &= 0 &&\text{ on }\partial\Omega_\text{outflow}(t).
  \end{alignedat}
\end{equation}
These conditions are, in order, the imposed velocity profile on the inlet,
the slip conditions enforcing a tangential flow on the relevant parts of the boundary,
the no-slip condition on the solid obstacle,
and the stress-free outlet, often called the natural outflow condition.
The no-slip condition on the obstacle requires a match between the fluid velocity and the obstacle velocity:
in this case, the obstacle is rigid and the mesh is not allowed to slide on its boundary,
so that the no-slip amounts to match the fluid velocity with the mesh velocity on the obstacle.

\begin{figure}
  \centering
  \includegraphics[width=\linewidth]{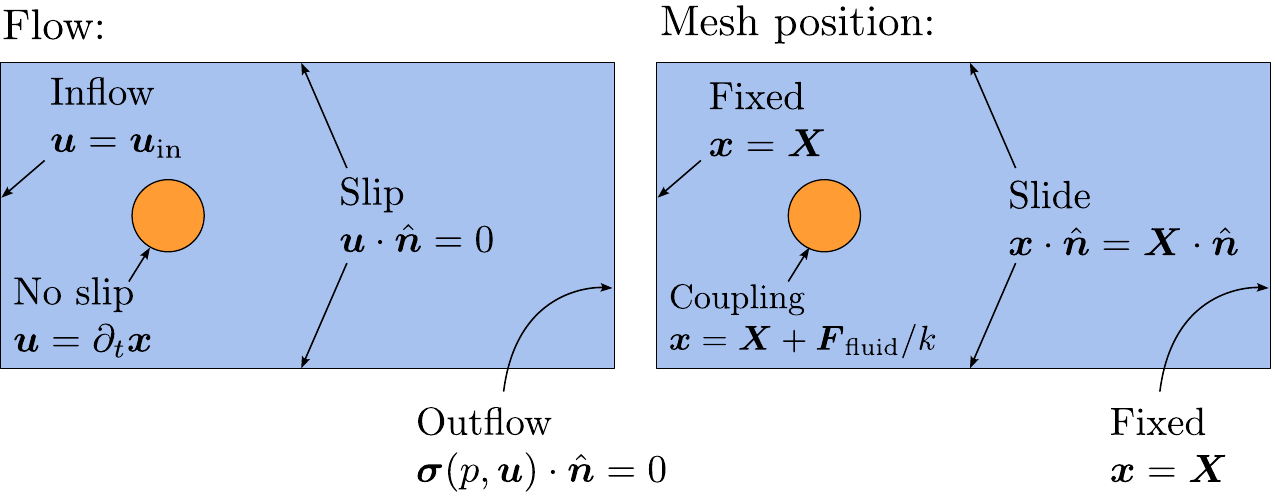}
  \caption{Boundary conditions for the flow and mesh position equations.}
  \label{fig:bc}
\end{figure}

For the linear elastic analogy, the mesh is fixed on the inlet and outlet,
but is free to slide on the other outer boundaries:
\begin{equation}
  \label{eq:psBC}
\begin{alignedat}{2}
  \bfx(\bfX) &= \bfX ~~~&&\text{ on }\partial\Omega_\text{fixed} = \partial\Omega_\text{inlet} \cup \partial\Omega_\text{outflow},\\
  \bfx(\bfX) \cdot \hatn &= \bfX \cdot \hatn &&\text{ on }\partial\Omega_\text{slide} = \partial\Omega_\text{slip},\\
  \end{alignedat}
\end{equation}
where $\hatn$ is the unit outward normal vector to the boundary.
These boundary conditions are summarized in \Cref{fig:bc}.
The Dirichlet boundary condition on the rigid, massless obstacle is obtained from the
solid's momentum balance \cref{eq:solid_balance}, which degenerates
into a force balance at its center of mass $\bfx_c$.
As the points on $\Gamma_t$ remain at constant distance from the center of the cylinder,
this yields the boundary condition:
\begin{equation}
  \label{eq:bc_solid}
  \bfx = \bfX + \frac{1}{k}\bfF_{\text{fluid}}(p, \bfu) ~~~\text{ on }\partial\Omega_\text{no slip} = \Gamma_t,
\end{equation}
which couples the mesh movement to the flow.
Along the $z-$component, the cylinder is fixed, and this boundary condition reduces to
$\bfx = \bfX$.

\section{Numerical scheme}
\label{sec:numerical_scheme}

The fluid-structure interaction system \cref{eq:system} is solved with the method of lines \cite{ern2004theory},
using a standard, second-order backward differentiation formula (BDF2) implicit scheme in time,
and a monolithic finite element formulation in space.
In this formulation, the velocity, pressure, forces and mesh movement are solutions of a single
system,
which requires
evaluating the fluid forces on the solid without postprocessing.
This can be done using the Lagrange multiplier framework,
in which the fluid forces result from the weak imposition
of the no-slip boundary condition on the solid.
In this context, the no-slip condition
 $\bfu = \partial_t \bfx$
is enforced as a constraint, with a Lagrange multiplier $\bflambda$ defined on $\Gamma_t := \partial\Omega_{\text{no slip}}$ only.
The multiplier is then interpreted as the fluid force density on $\Gamma_t$:
\begin{equation}
  \bfF_{\text{fluid}} = - \intgamt{\bflambda},
\end{equation}
as recalled hereafter.
The rest of this section details the weak formulation and function spaces considered, briefly presents the nonlinear solver and the additional terms of the Jacobian matrix arising from the moving grid, and discusses the choice of finite element spaces.

\subsection{Weak formulation}
\label{sec:weak_form}
To derive the weak formulation of \cref{eq:system},
the Navier-Stokes equations, augmented with the Lagrange multiplier terms,
are multiplied by test functions $\hatbfu$ and $q$
and integrated over $\Omega_t$,
and the pseudo-solid equation is multiplied by a test function $\hatbfx$
and integrated over $\Omega_0$.
The stress tensor divergence terms are integrated by parts to lower the derivative
order.
Lastly, the no-slip constraint $\bfu - \partial_t\bfx = 0$ becomes an equation of the system and is tested
against functions $\hatbflambda$.
This yields the problem of finding the quadruplet
$U := (\bfu, p, \bfx, \bflambda) \in \ucal := \vcal \times \qcal \times \xcal \times \lcal$
such that for all $(\hatbfu, q, \hatbfx, \hatbflambda) \in \ucal$ and all $t \in (0, T)$, we have:
\begin{equation}
  \label{eq:weak_system}
  \begin{alignedat}{2}
    \intomt{\hatbfu \cdot \rho \partial_t{\bfu}}
    +
    \intomt{\hatbfu \cdot \rho \left(\left(\bfu - \partial_t{\bfx}\right)\cdot \nabla\right)\bfu}
    ~~~~~~~
    &
    \\
    +
    \intomt{\nabla \hatbfu : \bfsigma_f(p, \bfu)}
    - \intdomt{\hatbfu \cdot (\bfsigma_f \cdot \hatn)}&
    \\
    - \intgamt{\hatbfu \cdot \bflambda}
    &= 0 ~~~~~~ &&\forall \hatbfu \in \vcal,
    \\
    -\intomt{q(\nabla \cdot \bfu)} &= 0, && \forall q \in \qcal,
    \\
    \intomzero{\nabla_{\bfX}\hatbfx : \bfsigma_s(\bfx)}
    -
    \intdomzero{\hatbfx \cdot (\bfsigma_s \cdot \hatn)}
    &= 0, && \forall \hatbfx \in \xcal,
    \\
    - \intgamt{\hatbflambda \cdot (\bfu - \partial_t\bfx)} &= 0, && \forall \hatbflambda \in \lcal.
  \end{alignedat}
\end{equation}
The integrals on the current
configuration $\Omega_t$ depend on the unknown position field $\bfx$,
whereas integrals on $\Omega_0$ are computed with respect to fixed coordinates.
These integrals in \eqref{eq:weak_system} are well-defined for $\bfu \in (H^1(\Omega_t))^d$,
$p \in L^2(\Omega_t)$, $\bfx \in (H^1(\Omega_0))^d$
and $\bflambda \in (H^{-1/2}(\Gamma_t))^d$,
and we define the function spaces:
  \begin{align}
    \vcal &= \left\{ \hatbfu \in (H^1(\Omega_t))^d ~ | ~ \hatbfu = 0 \text{ on }\partial\Omega_\text{inlet}, \hatbfu \cdot \hatn = 0 \text{ on }\partial\Omega_\text{slip}\right\},\\
    \qcal &= L^2(\Omega_t) ~/ ~\R,\\
    \xcal &= \left\{ \hatbfx \in (H^1(\Omega_0))^d ~ | ~ \hatbfx = 0 \text{ on }\partial\Omega_\text{fixed}, \hatbfx \cdot \hatn = 0 \text{ on }\partial\Omega_\text{slide} \right\},\\
    \lcal &= (H^{-1/2}(\Gamma_t))^d.
  \end{align}
These spaces are such that the boundary integrals on
$\partial\Omega_t /\Gamma_t$ and on $\partial\Omega_0$ vanish.
The remaining volume integral in the pseudo-solid solid equation can be written
in full, yielding
\begin{equation}
  \intomzero{\left(\lambda_s(\nabla_{\bfX} \cdot \hatbfx)(\nabla_{\bfX} \cdot \bfx - d) + 2\mu_s \nabla_{\bfX}\hatbfx : \bfvarepsilon(\bfx)\right)} = 0
\end{equation}
In the fluid momentum equation, the stress vector $\bfsigma_f \cdot \hatn$
in the boundary integral on $\Gamma_t$ has been replaced by $\bflambda$,
as is customary in the Lagrange multiplier approach.
Thus, $\bflambda$ is identified to the stress vector (i.e., the force density)
on the fluid, which is up to sign the fluid force density on the obstacle
(note that $\hatn$ is the outward unit normal to the \emph{fluid} domain $\Omega_t$,
not to the solid domain),
allowing an implicit computation of the fluid forces.
These forces are coupled to the position field on $\Gamma_t$ through
the boundary condition \cref{eq:bc_solid}, which is rewritten as:
\begin{equation}
  \label{eq:lambda_coupling}
  \bfx = \bfX - \frac{1}{k} \intgamt{\bflambda}.
\end{equation}

\subsection{Nonlinear solver and Jacobian matrix}
\label{sec:jacobian}
To write the weak formulation \eqref{eq:weak_system} in a more compact form,
we introduce the standard shorthand notations $(f,g)_\Omega := \intom{f \cdot g}$ on interior domains
and $\langle f, g \rangle_\Gamma := \intgam{f \cdot g}$ on boundaries,
where the dot is either a product of scalar fields, the dot product of vector fields
or the double contraction of 2-tensor fields.
One can then define a nonlinear functional $\NL(U)$ for the considered function spaces by
\begin{equation}
  \label{eq:nonlinear_functional}
\begin{alignedat}{1}
\NL(U)
&:= \left(\hatbfu, \rho \left[\partial_t \bfu + ((\bfu - \partial_t \bfx)\cdot \nabla) \bfu\right]\right)_\Omegat
+ \left(\nabla\hatbfu, \bfsigma_f(p, \bfu)\right)_\Omegat
\\[0.4em]
- &\left< \hatbfu, \bflambda\right>_\Gammat
- (q, \nabla \cdot \bfu)_\Omegat
+ (\nabla_{\bfX} \hatbfx, \bfsigma_s(\bfx))_{\Omega_0}
- \langle\hatbflambda, \bfu - \partial_t \bfx \rangle_\Gammat.
\end{alignedat}
\end{equation}
The weak formulation then amounts to find $U \in \ucal$
such that $\NL(U) = 0$ for all $V \in \ucal$.
This nonlinear problem is solved with a standard Newton-Raphson method.
The $m$-th Newton-Raphson iteration consists in solving for the
increment $\Delta U_{m}$ the linearized equation:
\begin{equation}
  \jcal(U_m) \Delta U_{m} = -\NL(U_m).
\end{equation}
In this expression, $\jcal(U)$ is the linearization of the functional $\NL$ (i.e., its Fréchet derivative),
and the term $\jcal(U) \Delta U$ is understood as the Gateaux derivative of $\NL$
in the direction $\Delta U$.
Each "column" of this "matrix" can be obtained by considering admissible variations
of the solution vector, e.g., of the form $\delta U = (\delta \bfu, 0, 0, 0)$ or $(0, 0, \delta \bfx, 0)$,
measuring the variation of $\NL$ when the velocity or position varies, respectively.
For these variations, the Gateaux derivative writes:
\begin{equation}
  \delta\NL(U, \delta U) := \jcal(U)\delta U = \frac{d}{d\epsilon}\NL(U + \epsilon \delta U)\biggr\vert_{\epsilon = 0}.
\end{equation}
For instance, for the part of $\NL$ associated with the convective term of the Navier-Stokes equations (here, without ALE correction),
the variation w.r.t. velocity is well-known and writes:
\begin{equation}
  \begin{aligned}
    \delta\NL(U, \delta U)_\text{conv, non-ALE}
      &= \frac{d}{d\epsilon}(\hatbfu, \rho [((\bfu + \epsilon \delta \bfu) \cdot \nabla)(\bfu + \epsilon \delta \bfu)])_\Omegat\biggr\vert_{\epsilon = 0}\\
      &= (\hatbfu, \rho[(\delta \bfu \cdot \nabla) \bfu + (\bfu \cdot \nabla) \delta \bfu])_\Omegat.
  \end{aligned}
\end{equation}
On fixed grids, these computations are straightforward as the support of the integrals are independent
of the resolved fields.
Here, the Navier-Stokes equations are solved on a moving grid attached to the
position $\bfx$, and we must account for the variations of the integrals in $\NL$
due to variations $\delta U = (0,0,\delta \bfx,0)$.
As a simple example, consider the linear functional $\ical$ measuring the area
of a mesh element $K_t$, written on the reference finite element
$\hat{K}$ with coordinates $\bfxi$:
\begin{equation}
  \ical := \int_{K_t} 1\,d\bfx = \int_{\hat{K}} \left|\frac{\partial \bfx}{\partial \bfxi}\right|\,d\bfxi = \int_{\hat{K}} \det J \,d\bfxi,
\end{equation}
where $J$ denotes the Jacobian matrix of the reference-to-physical transformation $T_t : \hat{K} \to K_t$.
This functional has nontrivial variation w.r.t. $\bfx$ through $J$, and following Jacobi's formula, we write:
\begin{equation}
  \begin{aligned}
    \delta \ical(U, \delta \bfx) = \frac{d}{d\epsilon}\ical(U + \epsilon \,\delta\bfx)\biggr\vert_{\epsilon = 0}
    &= \frac{d}{d\epsilon} \int_{K_t} 1\,d(\bfx + \epsilon \delta \bfx)\biggr\vert_{\epsilon = 0}\\
    &= \frac{d}{d\epsilon} \int_{\hat{K}} \left|\frac{\partial (\bfx + \epsilon \delta \bfx)}{\partial \bfxi}\right|\,d\bfxi\biggr\vert_{\epsilon = 0}\\
    &= \int_{\hat{K}} \det J \trace\left(J^{-1}\frac{\partial (\delta \bfx)}{\partial \bfxi} \right)\,d\bfxi\\
    &= \int_{\hat{K}} \det J \trace G\,d\bfxi,
  \end{aligned}
\end{equation}
with $G := J^{-1}\nabla_{\bfxi} (\delta \bfx) = \nabla_{\bfx}(\delta \bfx)$.
The variation of the terms in $\NL(U)$ follows from the product rule, while also accounting for the variations of the physical mesh gradients $\nabla_\bfx$ when $\bfx$ varies. For instance, the variation of the continuity integral writes:
\begin{equation}
  \begin{aligned}
    \delta_\bfx \left(\int_{K_t}q(\nabla \cdot \bfu)\,d\bfx\right)
    &=  \int_{K_t}q\, \delta_\bfx\big((\nabla_\bfx \cdot \bfu)\,d\bfx\big)\\
    &=  \int_{K_t}q\big(\delta_\bfx(\nabla_\bfx \cdot \bfu)\,d\bfx + (\nabla \cdot \bfu) \,\delta_\bfx(d\bfx) \big)\\
    &=  \int_{\khat}q\big(\trace(\nabla_\bfx\bfu \cdot G) + (\nabla \cdot \bfu) \trace G\big) \det J \,d\bfxi  \end{aligned}
\end{equation}
The overall Jacobian matrix of the fully coupled system has thus the following block structure:
\begin{align}
  \jcal =
  \begin{pmatrix}
    A_{\bfu\bfu} & A_{\bfu p} & A_{\bfu \bfx} & A_{\bfu \bflambda}\\
    A_{p \bfu} & 0 & A_{p\bfx} & 0\\
    0 & 0 & A_{\bfx\bfx} & 0\\
    A_{\bflambda \bfu} & 0 & A_{\bflambda \bfx} & 0\\
  \end{pmatrix},
\end{align}
and exhibits the zero diagonal blocks characteristic of both saddle-point
problems for $(\bfu, p)$ and $(\bfu, \bflambda)$.
Note that this block structure only accounts for the couplings originating
from the weak formulation, and thus does not include the force-position couplings ($A_{\bfx\bflambda}$ block)
due to the constraint \eqref{eq:lambda_coupling} on the cylinder.

\subsection{Spatial finite element discretization}
\label{sec:fem}
With the continuous framework introduced, we now detail the spatial
discretization.
Two conforming simplicial meshes $\tcal_0$ and $\tcal_t$, of $\Omegaz$ and $\Omegat$ respectively, are considered.
At each time step, the Lagrangian elastic analogy problem is solved on $\tcal_0$,
and the flow problem with ALE correction is solved on $\tcal_t$.
However, only one mesh is actually handled, as $\tcal_t$ is obtained
through the position field $\bfx(\bfX,t)$, as the current position of the mesh vertices of $\tcal_0$.

On $\tcal_0$, $\bfx_h$ denotes the finite element spatial discretization of $\bfx$,
and $\bfu_h, p_h, \bflambda_h$ denote the discretizations on $\tcal_t$.
We consider Lagrange finite elements, and denote by $\pcal_k$ the standard space of continuous and piecewise polynomial functions of degree $k$ on either $\tcal_0$ or $\tcal_t$,
either scalar or vector-valued.
LBB-stable Taylor-Hood $\pcal_2-\pcal_1$ mixed finite elements are used to discretize the velocity-pressure pair \cite{boffi2013mixed},
and the mesh position is linear, $\bfx_h \in \pcal_1$.

Using a Lagrange multiplier to enforce the no-slip condition on $\Gamma_t$
yields an additional saddle-point problem involving $\bfu$ and $\bflambda$,
for which defining a stable discrete pair is not trivial.
In two dimensions, Couture et al. \cite{couture2020new}
successfully computed VIV using an equal order quadratic pair for $(\bfu,\bflambda)$.
Although the pointwise behavior of the Lagrange multiplier is not discussed,
its integral (the total force on $\Gamma_t$) visually converges to the forces
computed using the stress vector.
Here, we make this choice as well and seek a piecewise quadratic $\bflambda_h$.
The main advantage is that the no-slip constraint is then enforced up to the precision of the linear solver (i.e., to machine precision with a direct solver),
whereas it converges with the mesh when using a linear approximation.
With manufactured solutions in two and three dimensions,
we observed for all tested cases that the integral of the Lagrange multiplier (i.e., the fluid forces)
converge with optimal rate, see \Cref{sec:mms_coupled}.
This is encouraging, as it is the integral of $\bflambda$, not its pointwise values, that drives the cylinder and mesh movement.
Additional tests (not presented here) suggest, however,
that pointwise convergence depends on the boundary conditions and on whether the
solid body has intersections with the remaining boundary of the fluid domain.

\subsection{Newton-Raphson iteration and semi-discrete scheme}
With the above notations, the discrete solution $U_h^{n+1}$ at each time step is obtained by solving the nonlinear problem $\NL(U_h^{n+1}) = 0$ in a fully coupled fashion,
which involves solving the following Newton-Raphson linear system for the increment $\delta U_h^{n+1}$ until convergence:
\begin{equation}
  \label{eq:NR_iteration}
  \jcal(U_{h,m}^{n+1})_{ij} \, (\delta U_{h,m}^{n+1})_j = -\NL_i(U_{h,m}^{n+1}),
\end{equation}
where $m \geq 0$ is the iterations counter.
The right-hand side of \eqref{eq:NR_iteration} writes:
\begin{equation}
  \label{eq:nonlinear_functional_discrete}
\begin{alignedat}{1}
-&\NL_i(U_{h,m}^{n+1}) = -\left(\utesti, \rho \left[(\partial_t\bfu_{h})_m^{n+1} + ((\bfu_{h,m}^{n+1} - (\partial_t \bfx_h)_m^{n+1})\cdot \nabla) \bfu_{h,m}^{n+1}\right]\right)_{\Omega_{t + \Delta t}}\\[0.4em]
&+ \left(\nabla\utesti, \bfsigma_f(p_{h,m}^{n+1}, \bfu_{h,m}^{n+1})\right)_{\Omega_{t + \Delta t}}
+ \left< \utesti, \bflambda_{h,m}^{n+1}\right>_{\Gamma_{t + \Delta t}}
+ (\ptesti, \nabla \cdot \bfu_{h,m}^{n+1})_{\Omega_{t + \Delta t}}
\\[0.4em]
&- (\nabla_{\bfX} \xtesti, \bfsigma_s(\bfx_{h,m}^{n+1}))_{\Omega_0}
+ \langle\ltesti, \bfu_{h,m}^{n+1} - (\partial_t \bfx)_{h,m}^{n+1} \rangle_{\Gamma_{t + \Delta t}}.
\end{alignedat}
\end{equation}

\section{Implementation details and parallel considerations}
\label{sec:implementation}

The resolution of \eqref{eq:nonlinear_functional_discrete} is done within an in-house solver
\footnote{\url{https://github.com/arthurbawin/fez}}
based
on the open-source finite
element framework \dealii \cite{arndt2025deal}.
Through extensive templating, \dealii allows for mostly dimension-independent
programming and provides wrappers to MPI facilities and linear algebra libraries (e.g., PETSc \cite{petsc-web-page}), considerably easing the development of a parallel solver prototype in both two and three dimensions.
In this section, we discuss some implementation details and choices,
namely the use of a solution-dependent mapping to represent the moving mesh,
three coupling strategies to enforce the force-position constraints,
and the use of MUMPS \cite{amestoy2000mumps} as solver for the linear system.

\subsection{Two mappings to handle the Lagrangian and ALE formulations}
\label{sec:mappings}
The pseudo-solid model is solved in a purely Lagrangian formulation on $\Omega_0$,
whereas the incompressible Navier-Stokes equations, as well as the no-slip constraint,
are solved on the deforming domain $\Omegat$, following the movement of the cylinder.
Handling these two computational domains is done with two mappings:
a fixed, linear mapping for $\Omega_0$, and a moving mapping described by the position field for $\Omegat$.
Within the \dealii library, these mappings are respectively implemented by the
\texttt{MappingFE} and \texttt{MappingFEField} class templates.
The latter is tied to the position part of the global solution vector $U$,
and updates the position of the mesh vertices as $U$ is modified, e.g., after each Newton iteration.
This allows for a fully coupled formulation, where the flow and mesh position
are updated consistently at each time step, and allows to reach optimal time convergence.

\subsection{Force-position coupling and parallel strategies}
\label{sec:coupling}
The fluid-structure coupling \eqref{eq:lambda_coupling} on the cylinder is an
affine relation between $\bfx$ and $\bflambda$, yielding an affine constraint
between their degrees of freedom:
\begin{equation}
  \label{eq:coupling_constraint}
  \bfx_i = \bfX_i - \frac{1}{k}\intgamt{\bflambda} = \bfX_i + \sum_{j} c_{j} \bflambda_j, \text{ with } c_{j} := -\frac{1}{k} \intgamt{\ltestj}.
\end{equation}
In \dealii, this kind of constraint is enforced with the \texttt{AffineConstraints} tools,
which enforce linear and affine constraints between degrees of freedom (also called "dofs" hereafter).
In \eqref{eq:coupling_constraint}, the Lagrange multiplier dofs
are so-called \emph{master}, unconstrained dofs, dictating the position of the cylinder to the constrained, or \emph{slave}, position dofs.
Enforcing the constraint at a given nonlinear iteration should not modify the values of $\bflambda$,
as it otherwise yields unwanted and counterintuitive retroaction of the pseudo-solid equation onto the
enforcement of the no-slip boundary condition on the cylinder
(for instance, the error on the no-slip enforcement then depends on the Lamé coefficients
of the mesh).
This distinction does not exist within the \texttt{AffineConstraints} framework,
where all dofs are considered equally constrained
\footnote{See for instance \dealii's documentation for affine constraints: \url{https://dealii.org/developer/doxygen/deal.II/classAffineConstraints.html}.}.
To circumvent this issue, the constraints \eqref{eq:coupling_constraint} are applied directly
to the linear system, modifying the rows, columns and right-hand side of the
constrained position dofs only.

\begin{figure}
  \centering
  \includegraphics[width=0.75\linewidth]{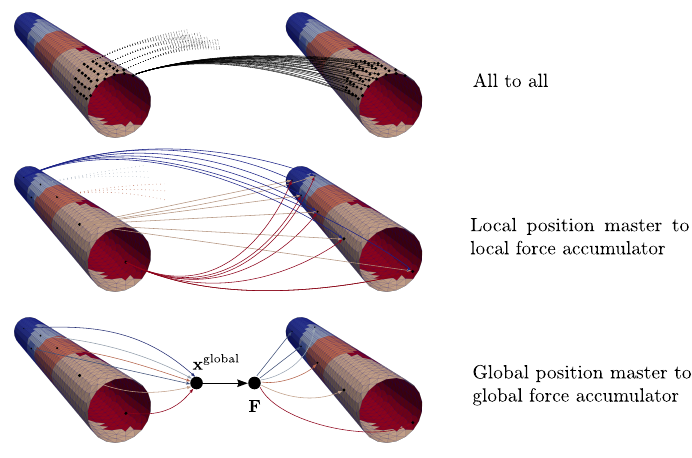}
  \caption{Coupling strategies}
  \label{fig:coupling_strategies}
\end{figure}

Since the cylinder is rigid, the force-position coupling \eqref{eq:coupling_constraint}
is nonlocal: the position dofs on the cylinder must move as one, and each dof must account
for the total resulting force on the cylinder,
including from remote mesh partitions.
Enforcing this constraint in parallel efficiently is challenging, as
$(i)$ owned and ghosted degrees of freedom must be constrained consistently,
to yield identical cylinder movement on arbitrary mesh partitions,
and $(ii)$
the number of additional couplings should remain modest,
to ensure an acceptable scaling as the number of MPI processes grows.
The former is an implementation challenge,
whereas the latter amounts to choosing an efficient coupling scheme between position and Lagrange multiplier dofs.
In this work, we considered the following three schemes,
defined by the use of either local or global so-called \emph{position masters} and \emph{force accumulators}.
These three strategies are illustrated on \Cref{fig:coupling_strategies}
for $N_\text{rank} = 6$ cylinder partitions,
and their performances and scaling are evaluated in \Cref{sec:results} in 2D and 3D.

\paragraph{All-to-all.} The bottom-line approach is a naive all-to-all scheme,
where all position dofs on the cylinder are explicitly coupled to the sum of all force contributions,
thus coupling them to all Lagrange multiplier degrees of freedom.
This can be schematized as:
\begin{equation}
  x^d_i \gets \sum_j c_j^d\lambda^d_j,
\end{equation}
with $d$ the space dimension, where $\gets$ denotes a one-way coupling.
Obviously, this scheme involves $N_\text{dofs}^\bfx \times N_\text{dofs}^\bflambda$ couplings
and a very large number of communications, and cannot scale
as the number of MPI processes (and
thus the number of partitions on the cylinder) increases.

\paragraph{Local position masters to local force accumulators.}
The number of couplings can be reduced, on the one hand, by constraining all local
position dofs on a partition to a single master per spatial dimension,
and, on the other hand, by accumulating the local force contributions into dedicated degrees of freedom.
On a partition with rank $r$ and with a chunk of the cylinder, a local position master degree of freedom for each
space dimension $x^{d, \text{loc}}_{r}$ is defined, and each other local position dof
is set to this master.
Similarly, local force accumulators are defined as the sum of local Lagrange multiplier dofs,
$F^{d, \text{loc}}_r := \sum_j c_j^d \lambda^d_{j,r}$, and the coupling can be written as:
\begin{equation}
  x^d_{i,r} \gets x^{d, \text{loc}}_{r} \gets \sum_{r = 1}^{N_\text{ranks}} F^{d, \text{loc}}_r \gets \sum_j c_j^d \lambda^d_{j,r}
\end{equation}
This scheme involves $d N_\text{ranks}^2$ couplings across partitions from each
local position master to each local force accumulator,
and additional $N_{\text{dofs},r}^\bfx + N_{\text{dofs},r}^\bflambda$ couplings
local to each partition.
In practice, the local position master dofs are chosen arbitrarily from among the local dofs, but force accumulators are additional degrees of freedom that must be added to the system.

\paragraph{Global position masters to global force accumulators.}
To further reduce the number of couplings, a single global position master dof and
force accumulator can be defined for each space dimension, for instance
on the partition with lowest rank from among those owning a chunk of the cylinder.
For each space dimension $d$, the coupling scheme becomes:
\begin{equation}
  x_{i,r}
  \overset{N_{\text{dofs},r}^\bfx}{\longleftarrow}
  x^{\text{loc}}_{r}
  \overset{N_\text{ranks}}{\gets}
  x^{\text{glo}}
  \overset{1}{\gets}
  F^{\text{glo}}
  \overset{N_\text{ranks}}{\gets}
  \sum_{r = 1}^{N_\text{ranks}} F^{\text{loc}}_r
  \overset{N_{\text{dofs},r}^\bflambda}{\gets}
  \sum_j c_j \lambda_{j,r},
\end{equation}
involving $2dN_\text{ranks}$ couplings across partitions.

\subsection{Linear solver}
\label{sec:direct}
For prototyping, the linear system at each Newton iteration is solved with the parallel
sparse direct solver MUMPS, interfaced in \dealii through PETSc.
This solver is used for both 2D and 3D applications.
Parallel direct solvers are known not to scale well with the number of MPI processes,
especially in 3D, and require considerable memory resources.
This aspect is the limiting factor of our current method, and looking into efficient preconditioners
(e.g., geometric multigrid on structured quadrilateral or hexahedral meshes) to use with a GMRES method is the topic of upcoming work.

\section{Code verification and validation}
\label{sec:v_and_v}

This section presents verification and validation tests for the monolithic solver
in two and three dimensions. We first assess the implementation of the proposed solver
with the method of manufactured solutions (MMS) \cite{roache2002code}, and show
that the expected spatial and temporal convergence rates are obtained through systematic
mesh and time step refinement.
Validation against experimental data is not considered, as the proposed model is
an idealized massless solid body. Instead, we present standard validation benchmarks,
the flow around a cylinder in two dimensions and around a sphere in three dimensions.

\subsection{Verification with manufactured solutions}
\label{sec:verification}
We proceed as follows:
\begin{itemize}
  \item First, we consider the decoupled fluid and mesh problems, to show the correct implementation of both the elasticity equation and the Navier-Stokes system in an ALE setting augmented with the Lagrange multiplier equation. To this end, we consider a weakly enforced no-slip boundary condition on the cylinder, while the mesh is deformed independently of the fluid.
  \item Then, the fully coupled problem is considered with an appropriate choice of manufactured solution, with mesh movement coupled to the fluid forces.
\end{itemize}

To mimic the flow around a cylinder, the geometry of interest in two dimensions is the unit square $[0,1]^2$ with a circular hole of radius $R_0 = 0.15$,
and the box $[0,1]^2 \times [0, 0.5]$ traversed by a circular cylinder of same radius in three dimensions,
\Cref{fig:mms_geometry}.
The time interval is $[0,0.5]$ in both cases.
The fluid properties are set to the arbitrary values $\rho = 9.876, \mu_f = 1.234$,
and the Lamé coefficients of the mesh are set to $\lambda_s = \mu_s = 1$.

Convergence studies are performed with respect to both space and time discretizations,
on 8 meshes of increasing density in two dimensions, and on 6 meshes in three dimensions.
The time step is halved at each convergence step, starting with $\Delta t = 0.1$.
The following $L^\infty$ error norms in time
are considered:
\begin{equation}
\begin{alignedat}{1}
  e_\bfu
  &:= \max_{t_k} \vert \bfu(\bfx, t_k) - \bfu_h(\bfx, t_k) \vert_{H^1(\Omega_t)},\\
  e_p
  &:= \max_{t_k} \Vert p(\bfx, t_k) - p_h(\bfx, t_k) \Vert_{L^2(\Omega_t)},\\
  e_\bfx
  &:= \max_{t_k} \vert \bfx(\bfX, t_k) - \bfx_h(\bfX, t_k) \vert_{H^1(\Omega_0)},\\
  e_\bflambda
  &:= \max_{t_k} \vert \bflambda(\bfx, t_k) - \bflambda_h(\bfx, t_k) \vert_{L^2(\Gamma_t)},
\end{alignedat}
\end{equation}
as well as the maximum componentwise error on the total fluid force on the cylinder:
\begin{equation}
  e_{F_{i}}
  := \max_{t_k} \left\vert \left(\intgamt{\bflambda(\bfx,t_k)} - \intgamt{\bflambda_h(\bfx,t_k)}\right) \cdot \bfe_i\right\vert.
\end{equation}

\subsubsection{Manufactured solution for the decoupled systems}
\begin{figure}[hbtp]
  \includegraphics[width=\linewidth]{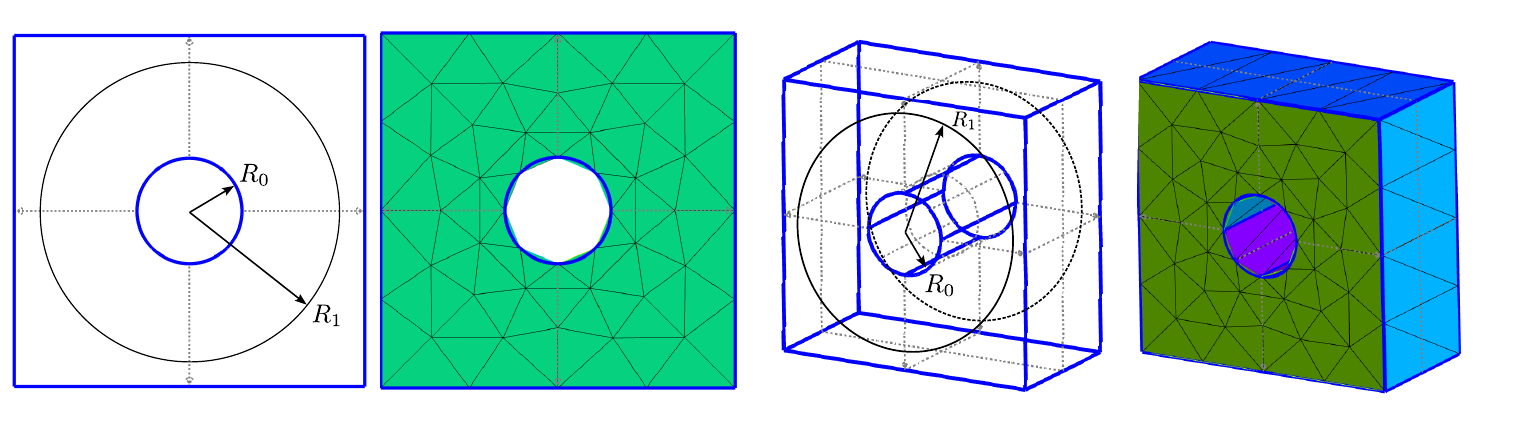}
  \caption{Geometries and initial meshes for the verification tests in two and three dimensions, and radius of influence of the manufactured mesh position.}
  \label{fig:mms_geometry}
\end{figure}

The manufactured solutions for the velocity, pressure, and Lagrange multiplier
are arbitrary, and appropriate source terms are added to each equation.
In two dimensions, the manufactured solution is:
\begin{equation}
  \begin{aligned}
    \bfu &= t^3 \left(\sin(\pi x)\cos(\pi y), \sin(\pi x)\sin(\pi y) \right),\\
    p &= t^2 \cos(\pi x)\cos(\pi y),\\
    \bflambda &= t^3 \left(\sin(\pi x)\sin(\pi y), \cos(\pi x)\cos(\pi y) \right),
  \end{aligned}
\end{equation}
For the three-dimensional case, the manufactured solution is:
\begin{equation}
  \begin{aligned}
    \bfu &= t^3 (\sin(\pi x)\cos(\pi y) \sin(\pi z),\\
    &\hspace{3cm}\cos(\pi x)\cos(\pi y)\cos(\pi z),\\
    &\hspace{6cm}\sin(\pi x)\sin(\pi y) \cos(\pi z)),\\
    p &= t^2 \cos(\pi x)\sin(\pi y)\cos(\pi z),\\
    \bflambda &= t^3 \left(x^3 + y^3, z^3, xyz \right),
  \end{aligned}
\end{equation}
In both cases, the manufactured position is chosen to deform the mesh while
keeping a rigid cylinder.
This is done with a smooth kernel function, and the prescribed position writes:
\begin{equation}
  \bfx(\bfX, t) = \bfX + f(t)\, \left(\bfX_{c, \text{final}} - \bfX_{c,0} \right) \psi(|\bfX - \bfX_c|),\\
\end{equation}
where $f(t) = t$ is the time dependence, $\bfX_{c, 0}$ and $\bfX_{c, \text{final}}$ are the initial and final locations of the cylinder respectively, and $\psi(|\bfX - \bfX_c|)$ is a bell-shaped kernel defined by
\begin{equation}
  \psi(r) := 1 - \varphi\left(\frac{r - R_0}{R_1-R_0}\right) ~~~ \text{with} ~~~ \varphi(s) := 6s^5 - 15s^4 + 10s^3,
\end{equation}
with $R_1 = 0.45$ the limit of the influence zone of the kernel, \Cref{fig:mms_geometry}.
Since $\varphi(0) = 0$ and $\varphi(1) = 1$,
this kernel satisfies $\psi(|\bfX_{\Gamma_t} - \bfX_c|) = \psi(R_0) = 1$,
so that all mesh vertices on the boundary of the cylinder are prescribed the same position.
Thus, this manufactured mesh position moves the cylindrical hole without deformation from its
initial to final location, and mimics the behavior of a rigid cylinder.

\begin{figure}[hbtp]
  \includegraphics[width=\linewidth]{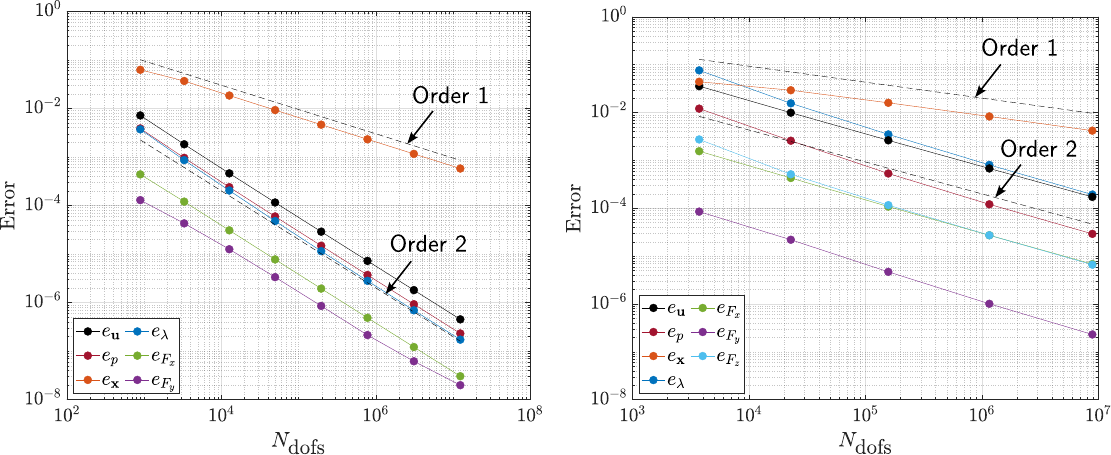}
  \caption{Convergence rates with manufactured solutions for the uncoupled case in two (left) and three (right) dimensions.}
  \label{fig:verification_decoupled}
\end{figure}

Convergence results for the decoupled tests and with a BDF2 time integration scheme
are shown in \Cref{fig:verification_decoupled}.
In both two and three dimensions,
the velocity, pressure and mesh position converge with an optimal rate:
order 2 for the velocity $H^1$-seminorm and for the pressure $L^2$-norm,
and order 1 for the position $H^1$-seminorm.
The Lagrange multiplier converges at order 2 in $L^2$-norm,
and the fluid forces are second-order accurate.
From these,
we conclude that neither the weak enforcement of the no-slip boundary condition,
nor the mesh movement, impact the convergence rates of the flow variables.

\subsubsection{Manufactured solution for the coupled system}
\label{sec:mms_coupled}
\begin{figure}
  \includegraphics[width=\linewidth]{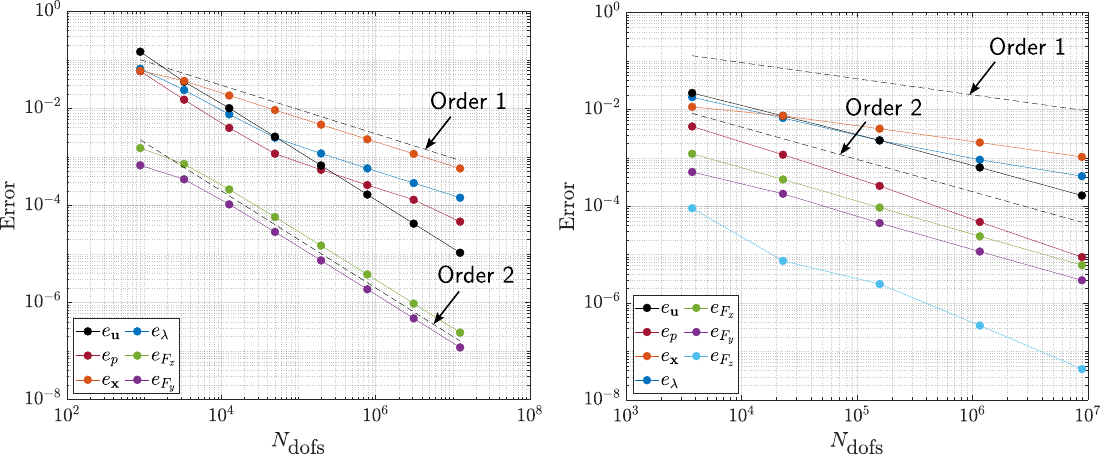}
  \caption{Convergence rates with manufactured solutions for the coupled case in two (left) and three (right) dimensions.}
  \label{fig:verification_coupled}
\end{figure}

We now assess the accuracy of the fluid-structure coupling.
Because it is delicate to add a source term to the constraint \cref{eq:lambda_coupling},
we consider instead a \emph{quasi}-manufactured solution.
This solution satisfies both \cref{eq:lambda_coupling}
and the no-slip boundary condition on the cylinder, so that no additional source
term is required to satisfy these constraints.
The manufactured solution in dimension~$d$~is:
\begin{equation}
  \begin{aligned}
    \bfu &= f'(t) \, \bfd \, \psi(|\bfx - \bfx_c(t)|),\\[1em]
    p &= -\frac{k\, f(t)}{C_d \pi R_0} \, \left(\bfd \cdot \frac{\bfx - \bfx_c(t)}{\Vert \bfx - \bfx_c(t) \Vert} \right) \, \psi(|\bfx - \bfx_c(t)|),\\[1em]
    \bfx &= \bfX + f(t)\, \bfd \, \psi(|\bfX - \bfX_c|), ~~~~~~~~~~~~~~ \bflambda = - \bfsigma(\bfu, p) \cdot \hatn,
  \end{aligned}
\end{equation}
where $\bfd := \left(\bfX_{c, \text{final}} - \bfX_{c,0} \right)$ is the total prescribed displacement, $f(t) = \sin(2\pi t)$, $\bfx_c(t) = \bfx(\bfX_{c,0}, t)$ is the current position of the center of the cylinder, $k$ is the spring stiffness, and where the constant $C_d = 1$ in two dimensions and $4R_0/3$ in three dimensions.
Since $\bfx|_{\Gamma_t} - \bfx_c(t) = \bfX|_{\Gamma_t} - \bfX_c $ because the cylinder is
rigid, the no-slip condition $\bfu|_{\Gamma_t} = \partial_t \bfx|_{\Gamma_t}$ is satisfied
exactly at all times.
The velocity gradient on the cylinder is zero due
to the fact that $\varphi'(0) = 0$, and the manufactured pressure is such that
the total force on the cylinder is:
\begin{equation}
  \begin{aligned}
    \bfF_{\Gamma_t}
    &= \intgamt{\bflambda}
    = \intgamt{- \bfsigma(\bfu, p) \cdot \hatn}
    = \intgamt{p\hatn}
    = k \, f(t) \, \bfd.
  \end{aligned}
\end{equation}
The fluid-structure coupling \cref{eq:lambda_coupling} then writes:
\begin{equation}
  \begin{aligned}
    \bfx|_{\Gamma_t}
    &= \bfX|_{\Gamma_t} - \frac{1}{k} \bfF_{\Gamma_t}
    = \bfX|_{\Gamma_t} + f(t) \,\bfd,
  \end{aligned}
\end{equation}
which is indeed satisfied on $\Gamma_t$, where $\psi = 1$.
Convergence results are shown in \Cref{fig:verification_coupled}.
Optimal convergence is obtained for the velocity and mesh position
in both two and three dimensions,
as well for the pressure in three dimensions.
In two dimensions, the pressure undergoes a reduced convergence but eventually goes back to second order
accuracy.
Although the convergence of the Lagrange multiplier falls to first order in both cases,
the hydrodynamic forces, driving the cylinder movement, converge at order 2,
which is satisfactory for the present application.

\subsection{Validation}
We consider here the unsteady incompressible flow around a two-dimensional cylinder at $Re = 200$,
and the steady-state flow around a sphere for Reynolds numbers ranging from 1 to 200.
For the flow around the cylinder, the quantities of interest are the force coefficients
(drag and lift) and the Strouhal number, characterizing the nondimensional vortex shedding:
\begin{equation}
  C_D = \displaystyle\frac{\left(-\intgamt{\bflambda}\right) \cdot \bfe_x}{\frac{1}{2}\rho U_\infty^2 A},
  ~~~~~
  C_L = \frac{\left(-\intgamt{\bflambda}\right) \cdot \bfe_y}{\frac{1}{2}\rho U_\infty^2 A},
  ~~~~~
  St = \frac{f_\text{s}D}{U_\infty},
\end{equation}
where $A = D$, and the vortex shedding frequency $f_s$ is computed from a fast Fourier transform on the periodic part of the
lift force signal.
For the flow around the sphere, the relevant indicator is the drag coefficient,
with $A = \pi D^2/4$.
In two dimensions, the considered benchmark and geometry is the same as in Mendes et al. \cite{mendes1999analysis},
and we use BDF2 time integration with constant time step $\Delta t = 5 \times 10^{-3}$ to simulate the interval $[0, 300]$.
The mesh contains 28,407 triangles.
In three dimensions, the geometry is a sphere with $D = 1$ in a rectangular box of dimensions $[22D, 10D, 10D]$,
with the center of the sphere located at $(6D, 5D, 5D)$, and the mesh contains 55,650 tetrahedra.
In both cases, we set $U_\infty = 1$, $\rho = 1$ and $\mu_f = \nu = Re^{-1}$.

\begin{figure}
  \centering
  \includegraphics[width=\linewidth]{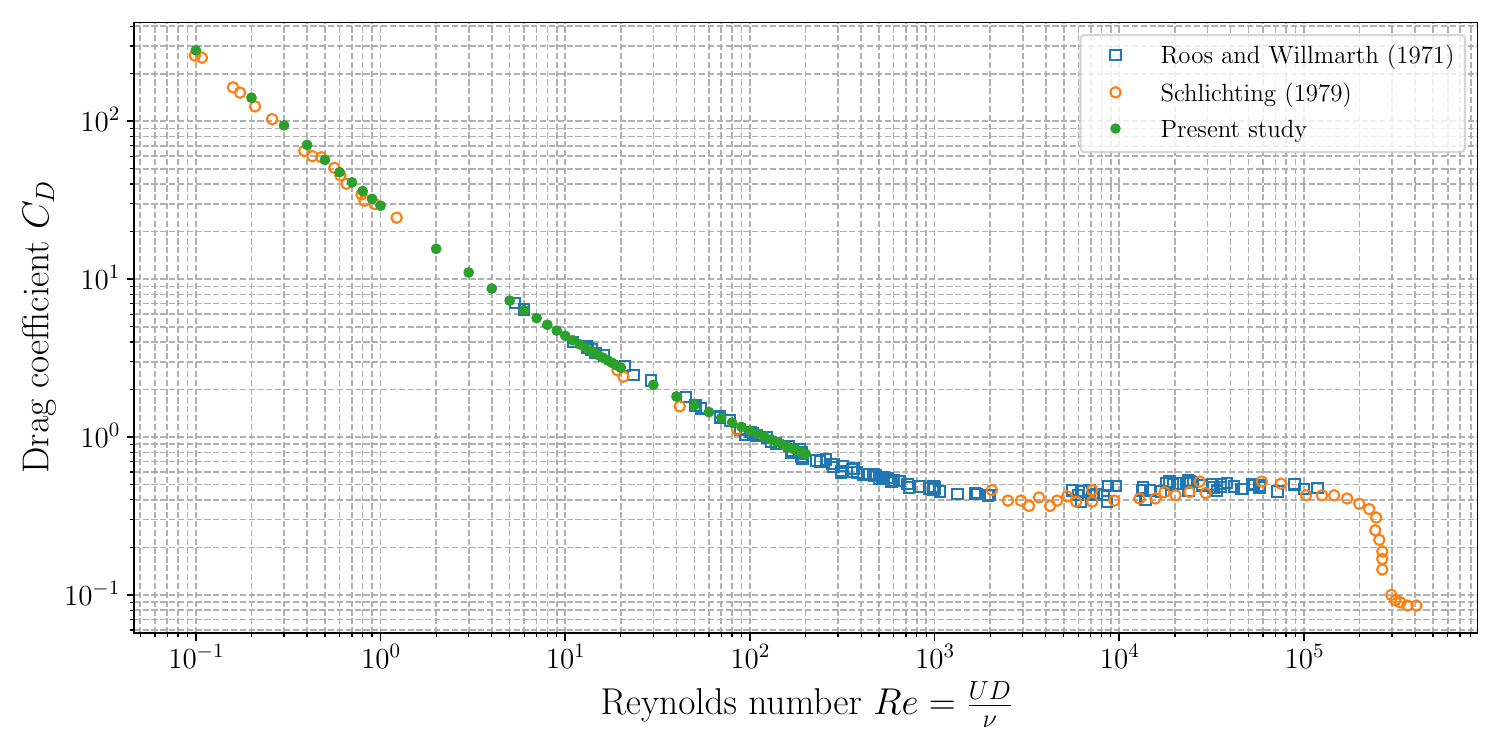}
  \caption{Steady incompressible flow around a sphere for $Re \in [1, 200]$: comparison of our predicted drag coefficient with those from the literature.}
  \label{fig:drag_sphere}
\end{figure}

Our numerical results are compared against values from the literature
in \Cref{tab:cylinder_benchmark} for the cylinder, including experimental data from Wieselsberger (presented by Roshko \cite{roshko1961experiments}, and \Cref{fig:drag_sphere} for the sphere.
In both cases, the comparison is excellent:
our cylinder indicators are within the expected ranges, and are particularly close
to the ones of Mendes et al. \cite{mendes1999analysis}, from whom the geometry is taken.
In three dimensions, our drag predictions are compared against the experimental data of
Roos et al. \cite{roos1971some}, and of Schlichting \cite{schlichting2016boundary}.
The agreement is excellent for the tested range of Reynolds numbers.
These tests demonstrate the accuracy of the proposed solver and its ability to predict external flows
when the no-slip boundary condition is enforced with a Lagrange multiplier.

\begin{table}[ht]
\centering
\begin{tabular}{lccc}
\hline
Reference & $C_D \pm \Delta C_D$ & $\pm C_L$ & $St$ \\
\hline
Wieselsberger (in \cite{roshko1961experiments}) & 1.38 & 0.77 & 0.18 \\
Lecointe and Piquet \cite{lecointe1984use} & 1.46 & -- & 0.194 \\
Braza, Chassaing and Minh \cite{braza1986numerical} & 1.38 & $\pm 0.77$ & 0.20 \\
Lecointe and Piquet \cite{lecointe1989flow} & $1.29 \pm 0.04$ & $\pm 0.60$ & 0.195 \\
Franke, Rodi and Schönung \cite{franke1990numerical} & 1.31 & $\pm 0.65$ & 0.194 \\
Zhang and Dalton \cite{zhang1997interaction} & $1.25 \pm 0.03$ & $\pm 0.54$ & 0.196 \\
Mendes and Branco \cite{mendes1999analysis} & $1.399 \pm 0.049$ & $\pm 0.726$ & 0.202 \\
Couture-Peck, Garon and Delfour \cite{couture2020new} & $1.3292 \pm 0.0349$ & $\pm 0.5882$ & 0.192 \\
\textbf{Present study} & $\mathbf{1.398 \pm 0.0497}$ & $\pm \textbf{0.723}$ & \textbf{0.199} \\
\hline
\end{tabular}
\caption{Unsteady incompressible flow around a circular cylinder at $Re = 200$: comparison of our predicted force coefficients and Strouhal number with those from the literature.}
\label{tab:cylinder_benchmark}
\end{table}

\section{Numerical results}
\label{sec:results}
We present simulations of vortex-induced vibrations (VIV) of a
massless cylinder in two and three dimensions.
The parametric space of interest is described by $m^* = 0$,
by $U_r \in [2, 12]$ and $Re \in [75,150]$ in two dimensions,
and by $U_r \in [ 2.5, 17.5]$ and $Re \in [100, 300]$ in three dimensions.
This allows for comparison with the two-dimensional studies of Yu et al. \cite{yu2018two}.

\begin{figure}
  \centering
  \includegraphics[width=\linewidth]{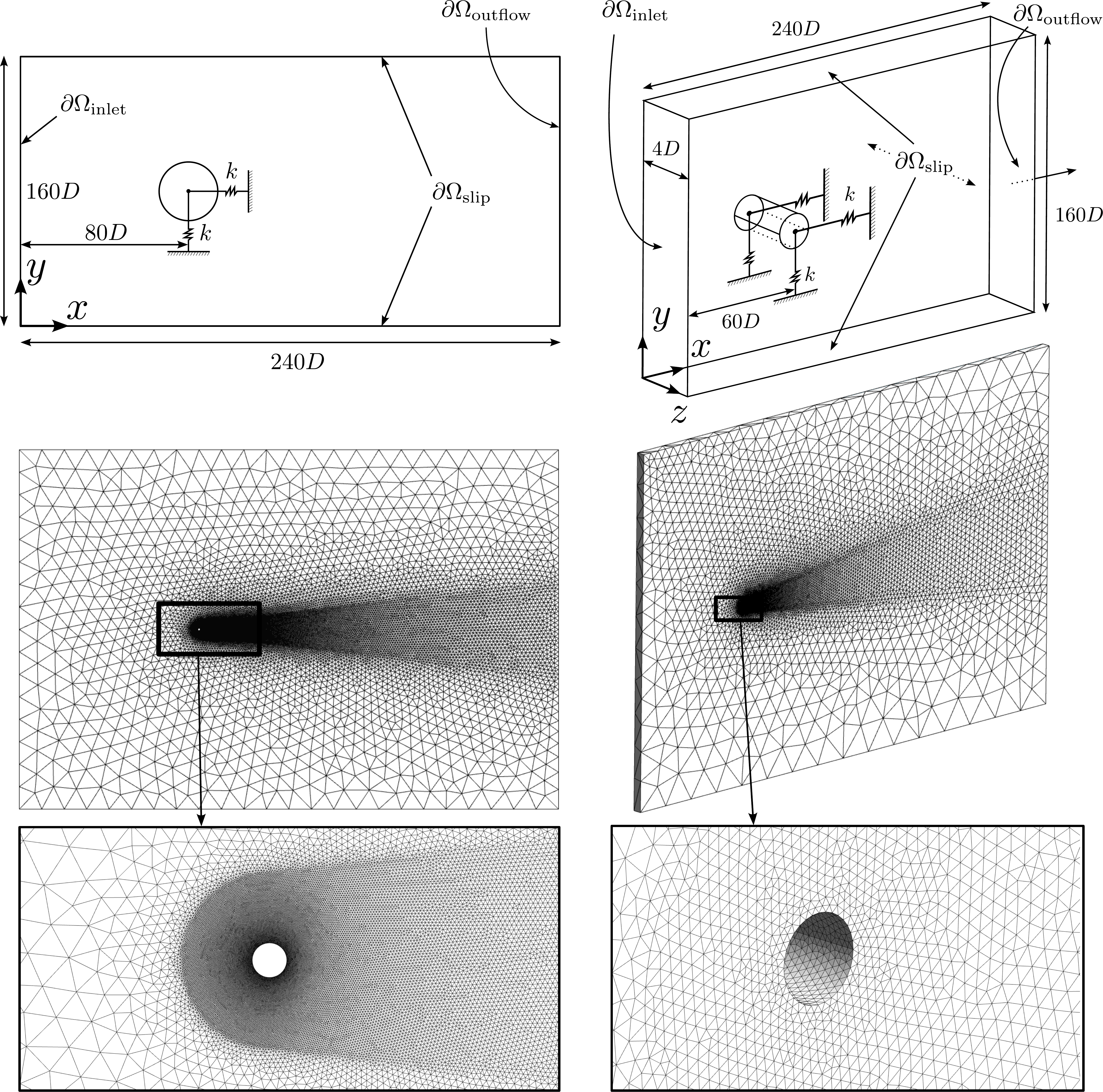}
  \caption{Geometry of the VIV problem (top, not to scale) and computational mesh for the 2D and 3D simulations (bottom).}
  \label{fig:viv_geometries}
\end{figure}

\paragraph{Geometry and mesh.}
The geometry in two dimensions is the one described in Yu et al. \cite{yu2018two}, \Cref{fig:viv_geometries},
and consists of a cylinder
of diameter $D = 1$ in a rectangular domain, mounted on two translational springs with constant
stiffness $k$. The dimensions of the box are $[240D, 160D]$, which is large enough to
reduce the influence of the boundaries.
The three-dimensional domain is an extrusion of the two-dimensional one along the $z$-direction
with length $4D$. The cylinder is free to move in the $xy$-plane, and is fixed in the $z$-direction.
Slip boundary conditions are applied on the front and back faces (i.e., the faces at $z = 0$ and $z = 4D$).
This reduced extrusion length is due to the use of a direct solver,
which limits the overall mesh size and currently prevents from simulating on wider 3D domains.
The wake and region around the cylinder is also less refined in 3D than in 2D, for this very reason.
The considered geometries and meshes are shown in \Cref{fig:viv_geometries}.

\paragraph{Time integration.} The simulations are run until a nondimensional time $t = 200$.
The starting time step is $\Delta t = 10^{-2}$, then adaptive time stepping is used
based on a target Courant-Friedrichs-Lewy (CFL) of 1, which acts as a simple convection error estimator.

\paragraph{Lamé coefficients.} The pseudo-solid elastic analogy does not prevent
inversion of the mesh elements during the simulation.
To preserve mesh positivity at all times even for reduced
spring stiffnesses and thus large mesh deformations, the Lamé coefficients are nonconstant across the
domain, and are set to the radial function
\begin{equation}
  \lambda_s(\bfX) = \mu_s(\bfX) = \left( \sqrt{(X-X_c)^2 + (Y-Y_c)^2} - (D/2)^2 \right)^{-2}.
\end{equation}
\noindent This distribution rigidifies the smaller elements near the cylinder,
letting large displacements be absorbed by the coarser elements farther in the domain.

\paragraph{Hardware.} All computations were run on the \emph{Fir} cluster from the Digital Research Alliance of
Canada.
Each node of \emph{Fir} consists of 2 AMD EPYC 9655 processors (Zen 5, 2.7 GHz), totalling
192 cores, and of 756 Gb of memory.
The tests presented hereafter were run on up to 8 compute nodes.

\paragraph{Results in two dimensions.}

\begin{figure}
  \centering
  \includegraphics[width=0.95\linewidth]{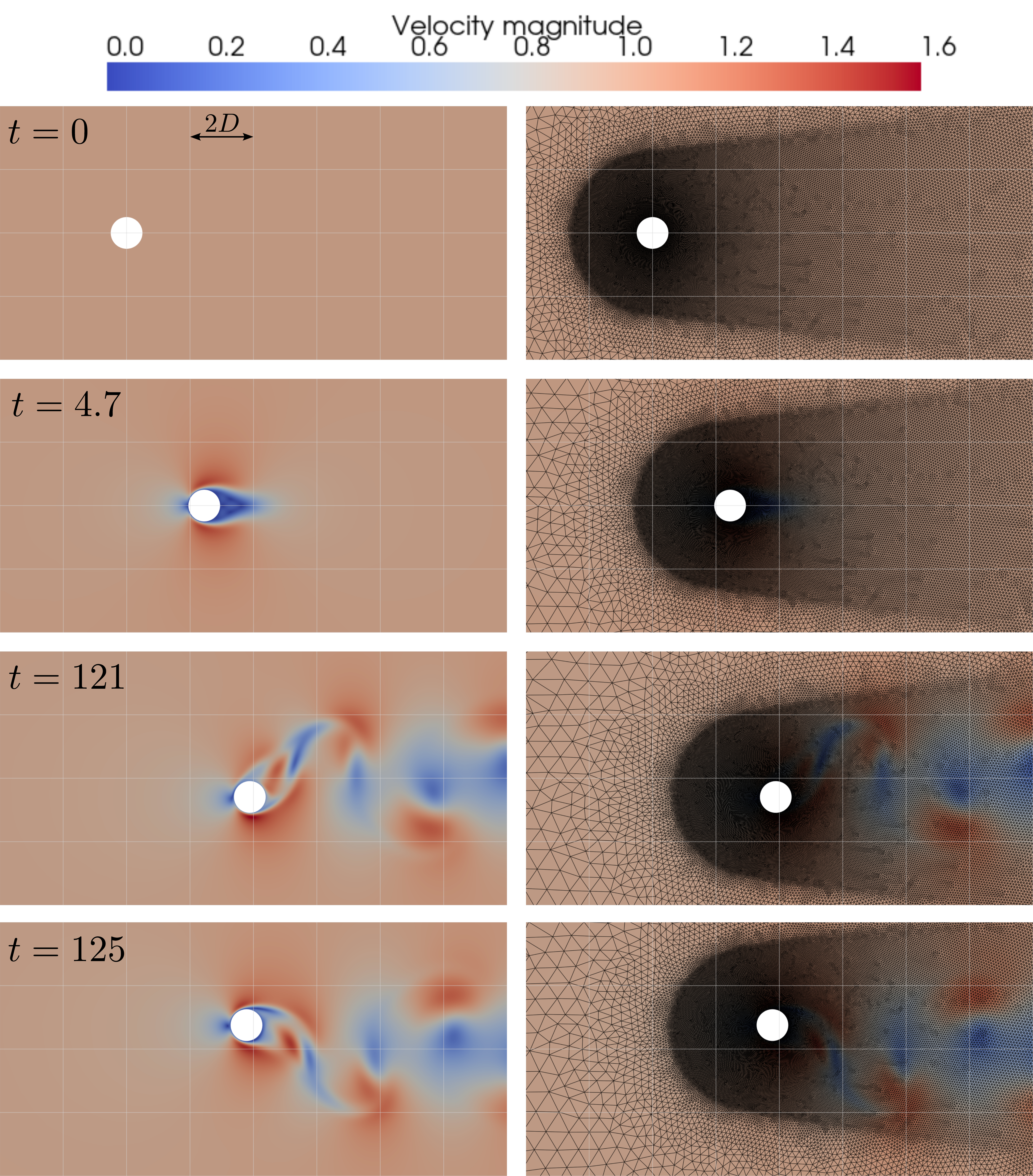}
  \caption{VIV in 2D: velocity magnitude and mesh for $Re = 125$ and $U_r = 10.5$.}
  \label{fig:velocity_2d}
\end{figure}

\begin{figure}
  \centering
  \includegraphics[width=\linewidth]{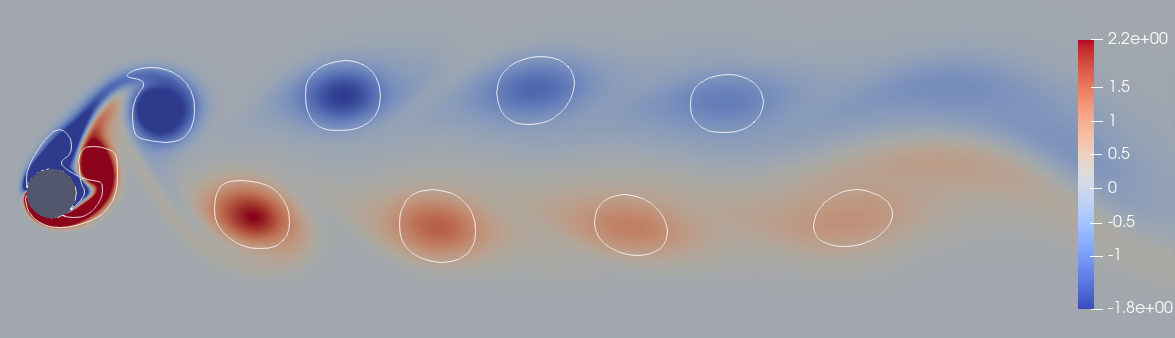}
  \includegraphics[width=\linewidth]{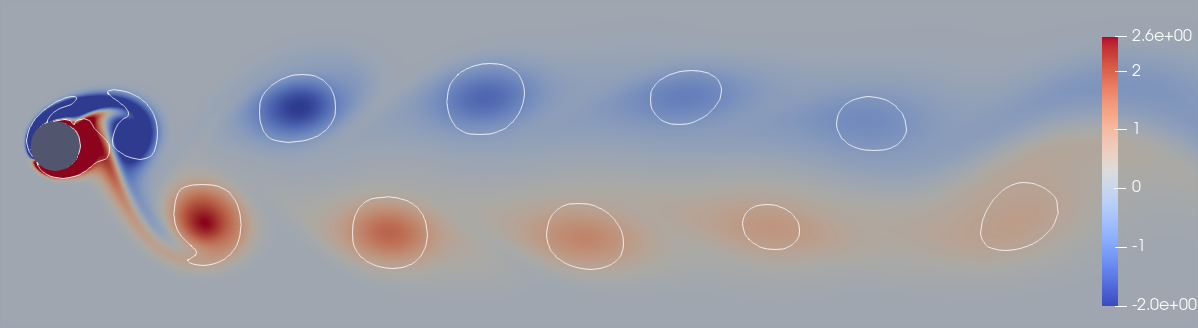}
  \caption{VIV in 2D: vorticity and isocontours 0.05 of the $Q$-criterion at $t = 115$ (top) and $t = 119$ (bottom) for $Re = 100$ and $U_r = 5$.}
  \label{fig:vorticity_2d}
\end{figure}

The evolution of the system and the velocity magnitude at times $t = 0, 4.7, 121$ and $125$ are shown in \Cref{fig:velocity_2d}
for $Re = 125$ and $U_r = 10.5$.
The wake develops from a uniform flow, while the mesh follows the displacement of the
cylinder without concentrating the deformation in the refined near-wall region.
Note that the field at $t = 0$ is the initial condition, a purely horizontal velocity
which does not satisfy the no-slip boundary condition. This condition is only satisfied after the
first time step, once the Lagrange multipliers have been computed.
At later times, larger deformations are absorbed by the coarser
elements away from the cylinder, which is the intended effect of the
nonuniform Lamé coefficients.
At $t = 4.7$, before the onset of the Von Karman streets,
the cylinder is pushed in the $x$-direction by an amplitude of about $2.5D$ (see grid).
The vortex streets are fully developed at $t = 125$, where the amplitude of the oscillations
in the $y$-direction is of the order of the cylinder diameter, and an order of $4D$ in the $x$-direction.
The right of \Cref{fig:velocity_2d} shows how the mesh follows and is deformed through the movement of the solid body.
Vorticity maps
are shown in \Cref{fig:vorticity_2d}, together with the isocontours 0.05 of the $Q$-criterion,
defined by $Q := (\Vert \boldsymbol{\Omega} \Vert^2 - \Vert \bfd \Vert^2)/2$,
where $\boldsymbol{\Omega}$ and $\bfd$ are the antisymmetric and symmetric parts of the velocity
gradient, respectively.
The two snapshots correspond to two
phases of the periodic regime. They show the alternating release of vortices
from the upper and lower sides of the cylinder, and the $Q$-criterion
isolines identify the coherent cores convected downstream.

\begin{figure}
  \centering
  \includegraphics[width=\linewidth]{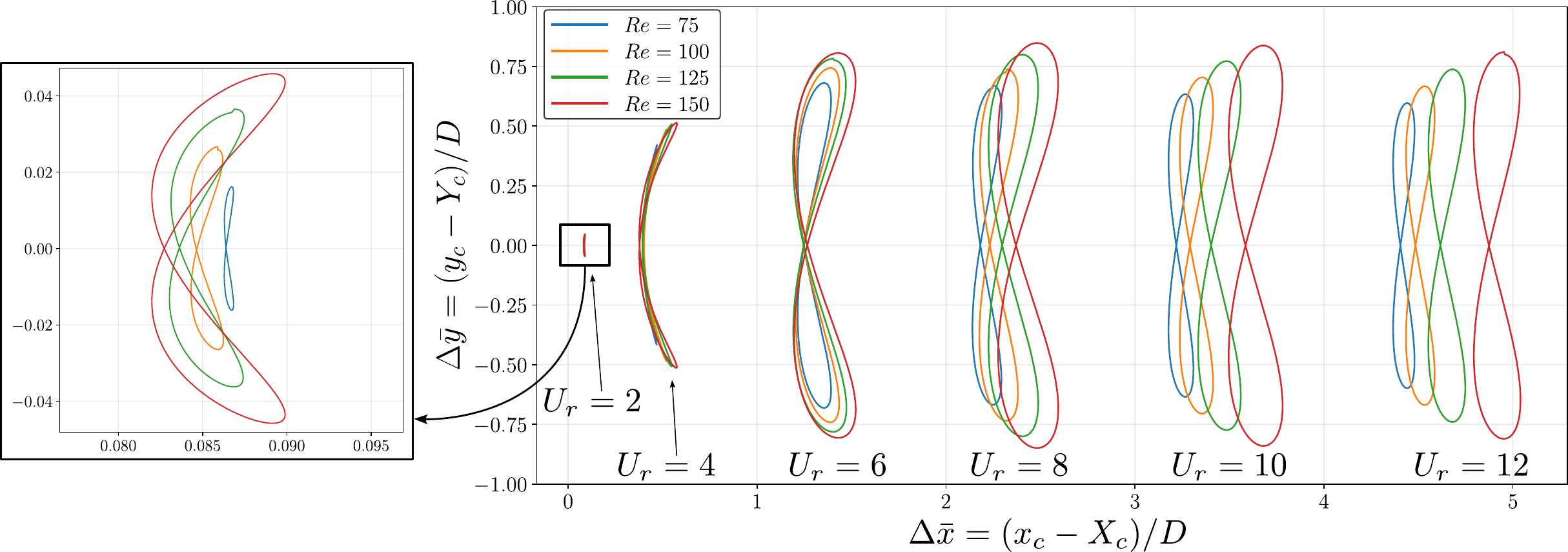}
  \caption{VIV in 2D: limit cycles of the trajectories of the center of the cylinder.}
  \label{fig:vid2d_trajectories}
\end{figure}

\paragraph{Trajectories and force coefficients.}
For the studied range of Reynolds numbers, the cylinder oscillations become
periodic after a transient period, after which its center describes
a lemniscate ("8" shape) in the $xy$-plane, \Cref{fig:vid2d_trajectories}.
All trajectories exhibit symmetric oscillations in the $y$-direction,
with maximum transverse oscillation $A_y := |y_c - Y_c| = 0.85D$ at $Re = 150$ and $U_r = 8$.
As the reduced velocity increases, the mean streamwise displacement
also increases, reflecting the smaller spring stiffness required to balance the
mean drag. The transverse amplitude is small at low $U_r$, grows rapidly over
the lock-in range, and remains large for the higher reduced velocities shown.
For a fixed $U_r$, increasing the Reynolds number generally increases the size
of the limit cycle, consistently with the
stronger vortex forcing
observations of Yu et al. \cite{yu2018two}.

\begin{figure}
  \centering
  \includegraphics[width=\linewidth]{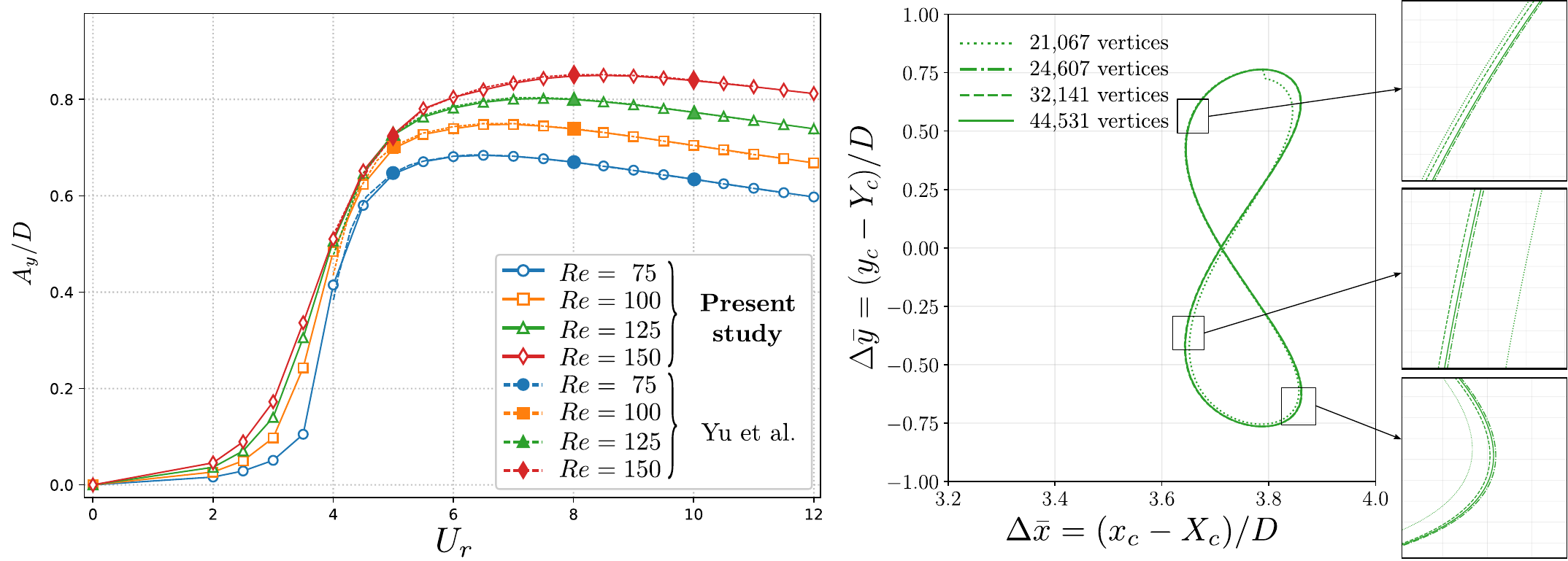}
  \caption{Left: comparison of the 2D oscillation amplitudes with reference results of Yu et al. \cite{yu2018two}.
  Right: mesh convergence of the limit cycle of the 2D trajectories for $Re = 125$ and $U_r = 10.5$.}
  \label{fig:present_study_vs_yu_2d_and_mesh_convergence}
\end{figure}

The amplitudes of the transverse oscillations obtained with the present
Lagrange multiplier formulation are nearly identical to the reference results of Yu et al. 
\cite{yu2018two}, using the method of reactions with governing equations written in a
moving frame, \Cref{fig:present_study_vs_yu_2d_and_mesh_convergence}.
This is highly encouraging, as both methods tackle the FSI problem in a quite
different way, but recover the same VIV response curves nonetheless.
The response curves show the expected low-amplitude branch at small reduced velocity,
followed by a sharp amplitude increase
around the onset of lock-in, and a broad
large-amplitude region featuring a monotonic increase of the response amplitude with the Reynolds number.

Mesh convergence of the oscillations in the periodic state is shown in
\Cref{fig:present_study_vs_yu_2d_and_mesh_convergence},
for four meshes of increasing density.
The comparison is performed at $Re = 125$ and $U_r = 10.5$, where the cylinder
undergoes large transverse oscillations and significant mesh deformation.
Although convergence is slightly oscillatory, the limit cycles obtained on the last three
meshes are very close to each other over the full trajectory.

\begin{figure}
  \centering
  \includegraphics[width=0.9\linewidth]{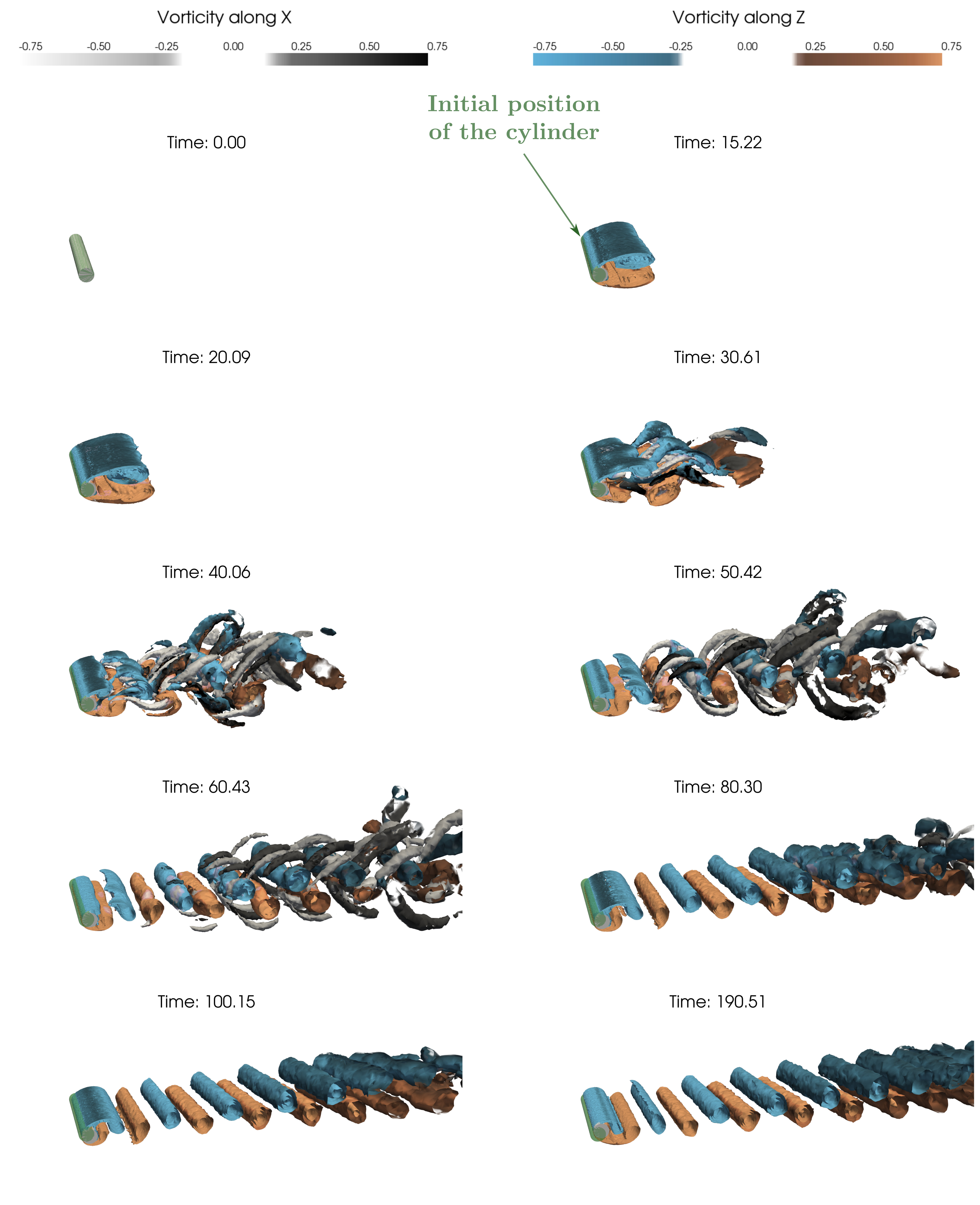}
  \caption{Vortex-induced vibrations of a massless cylinder in 3D: isosurfaces 0.01, 0.25 and 0.5 of the $Q$-criterion at $Re = 200$ and $U_r = 3$.}
  \label{fig:q_criterion_3d_200_3}
\end{figure}

\begin{figure}
  \centering
  \includegraphics[width=0.9\linewidth]{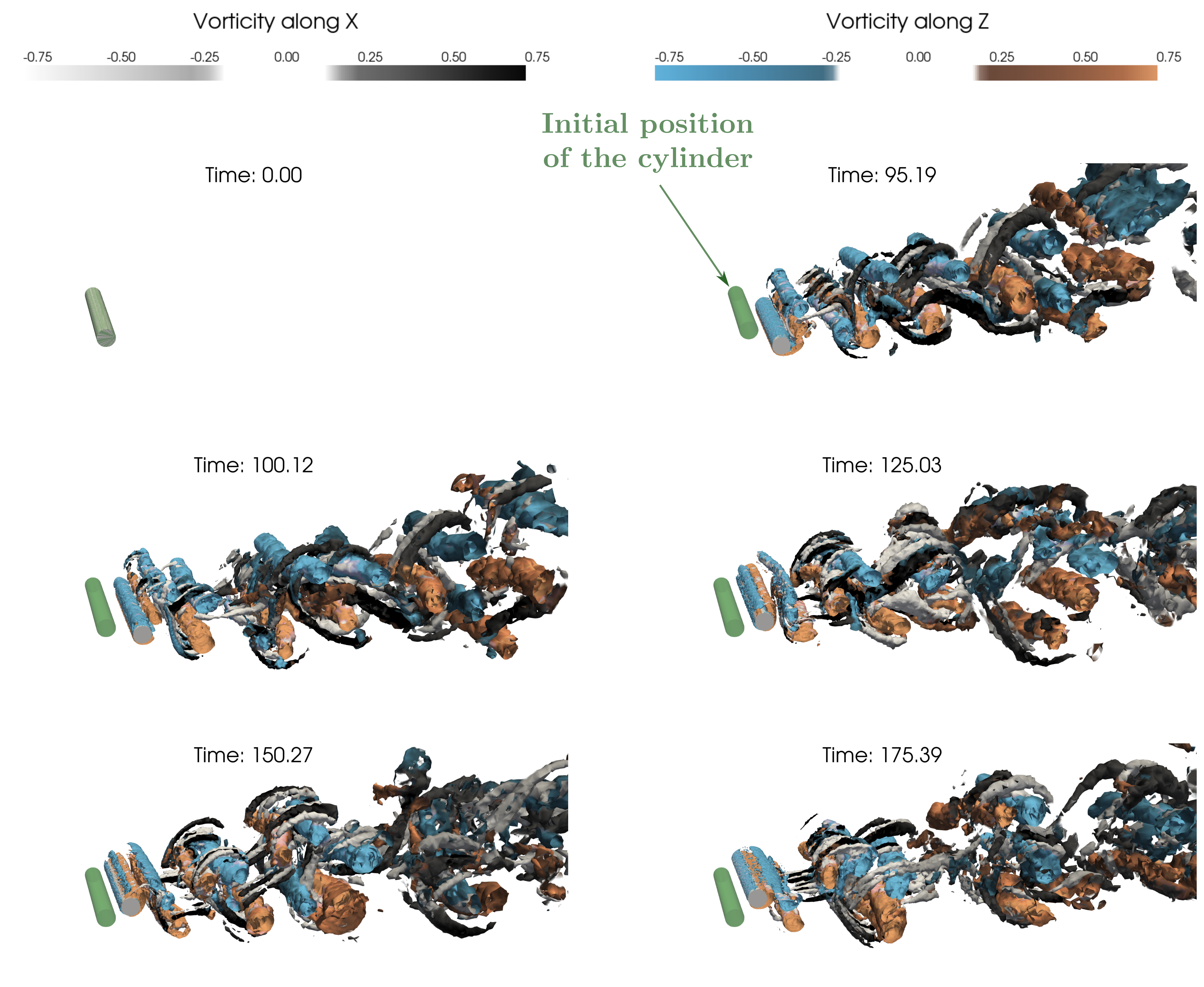}
  \caption{VIV in 3D: isosurfaces 0.01, 0.25 and 0.5 of the $Q$-criterion at $Re = 200$ and $U_r = 6$.}
  \label{fig:q_criterion_3d_200_6}
\end{figure}

\begin{figure}
  \centering
  \includegraphics[width=\linewidth]{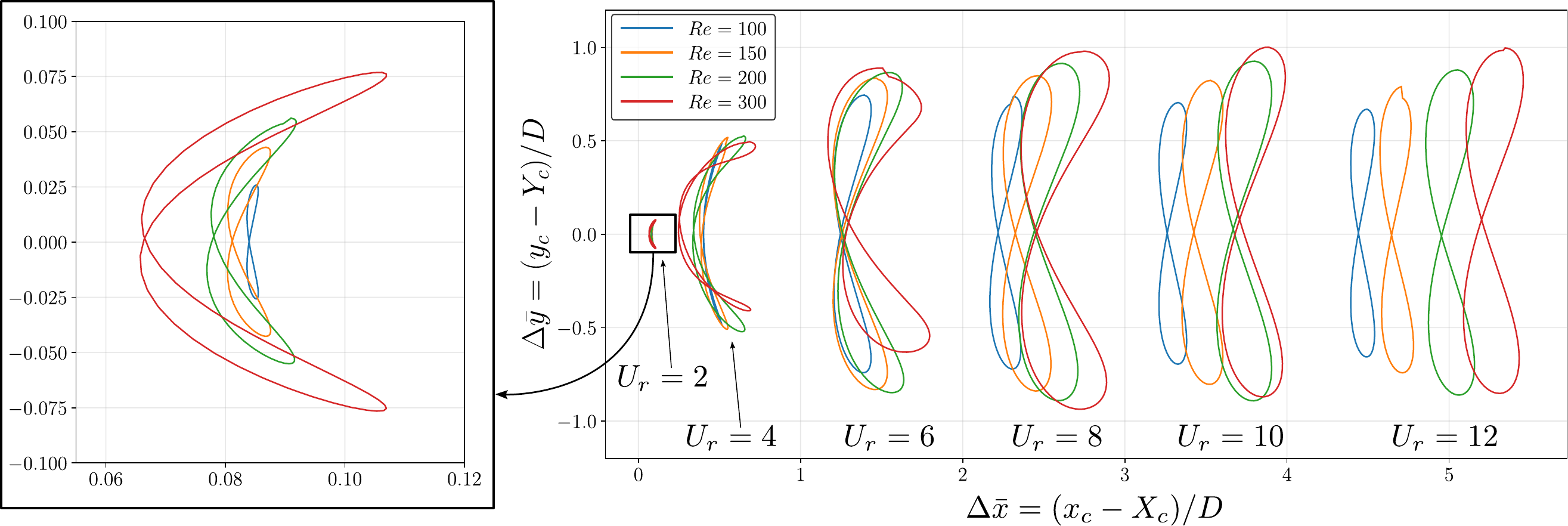}
  \caption{VIV in 3D: limit cycles of the trajectories of the center of the cylinder.
  The amplitudes are computed by averaging the last 10 periods.}
  \label{fig:trajectories_3d}
\end{figure}

\paragraph{Three-dimensional results.}
The evolution of the cylinder in three dimensions
and the isocontours 0.01, 0.25 and 0.5 of the $Q$-criterion are shown in
\Cref{fig:q_criterion_3d_200_3,fig:q_criterion_3d_200_6},
for combinations of Reynolds numbers 200 and 300, and reduced velocity $U_r = 3$ and 6.
On each figure, the initial position of the cylinder is highlighted in green,
to show the $xy$ cylinder motion.
For all configurations,
spanwise vortices are formed in the transient phase in addition to the streamwise vortices, and are convected to the outflow.
At $U_r = 3$, these vortices have disappeared when the periodic phase is reached,
and the periodic flows at both $Re = 200$ and $300$ are mostly two-dimensional.
At $U_r = 6$, the spanwise vortices are sustained throughout the simulation,
and the flows are fully three-dimensional.
The maximal vertical oscillation amplitude at this reduced velocity is 0.85$D$ (see also \Cref{fig:trajectories_3d}).

\begin{figure}
  \centering
  \includegraphics[width=0.65\linewidth]{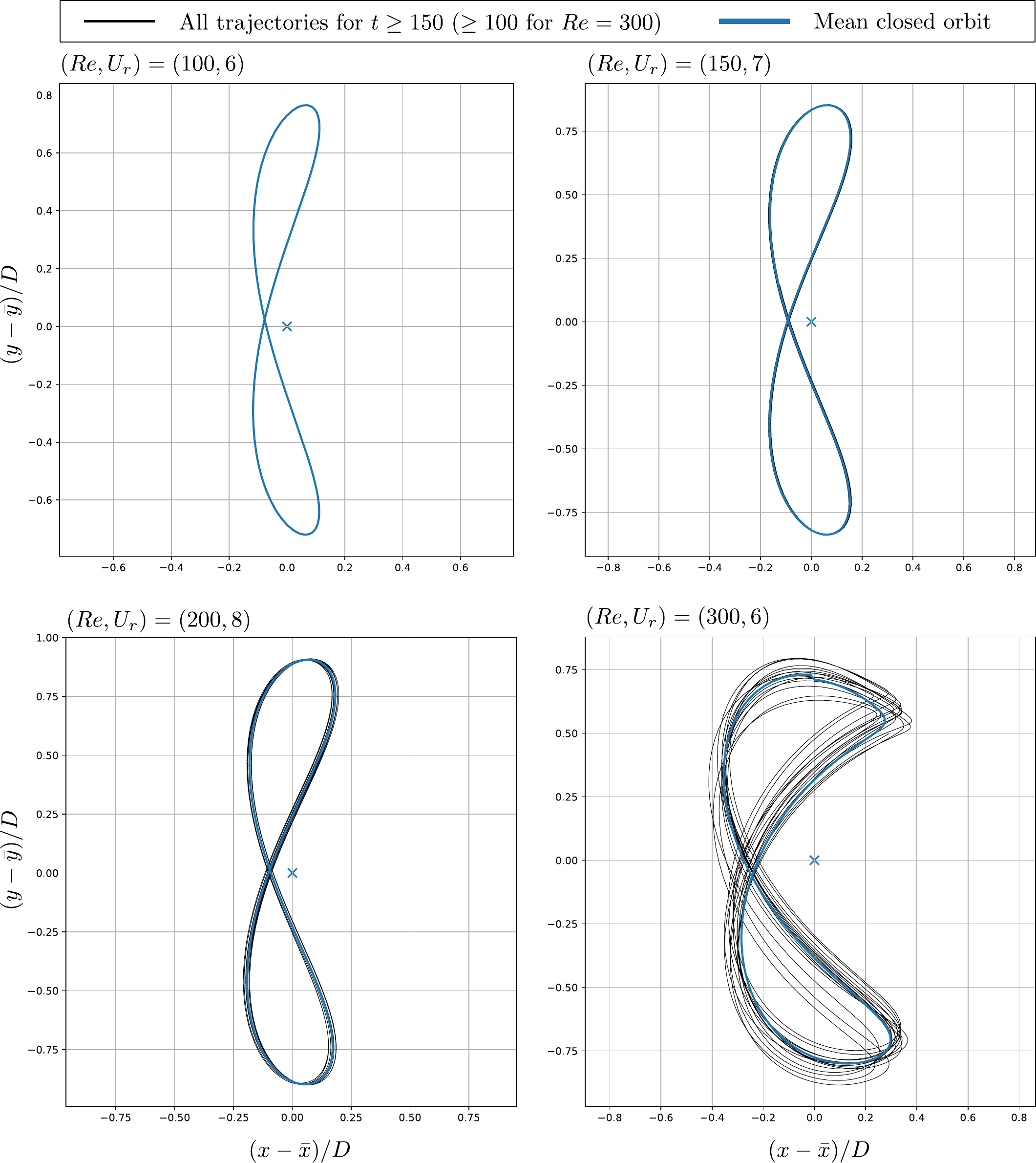}
  \caption{VIV in 3D: trajectories of the center of the cylinder.}
  \label{fig:all_trajectories_3d}
\end{figure}

The trajectories of the cylinder exhibit
the qualitative features observed in two dimensions, \Cref{fig:all_trajectories_3d}.
The locus described by its center is a lemniscate
with a dominant transverse oscillation and a streamwise
oscillation at twice the frequency.
At increasing $Re$ and reduced velocities, these trajectories are no longer periodic,
and at $(Re, U_r) = (300, 6)$, the system becomes chaotic: the center of the cylinder no longer describes a fixed orbit, and the center of its mean trajectory is no longer centered along the $x$-axis.
This asymmetry contributes to the peak oscillation,
which now consists of both a peak amplitude and an offset in the trajectory center.
Limit cycles of these trajectories
and force coefficients are shown in \Cref{fig:trajectories_3d}
at $Re \in [100, 300]$.
The mean streamwise
displacement increases with the reduced velocity, as the springs' stiffness is
decreased, and the transverse amplitude becomes large once the response enters
the lock-in range.

\paragraph{Scaling.}
Here, we compare the coupling strategies discussed in \Cref{sec:coupling},
and present timings and scalings with both full MPI and hybrid MPI-OpenMP configurations.
For these tests, we consider in 2D a medium and a fine mesh yielding 6 and 24 million degrees of freedom
respectively. In 3D, we consider a coarse, medium and fine mesh, yielding 0.6, 1.5 and 5.5 million dofs respectively.
The timings of interest are the assembly time (matrix and right-hand side (RHS)), which
includes the time taken to apply the force-position couplings, and the time spent solving the linear system
with MUMPS at each Newton-Raphson iteration.
For each test, the timings are the ones obtained after solving for the time interval $[0, 0.05]$, i.e.,
solving for the first 6 time steps with a BDF2 method with constant time step $\Delta t = 0.01$, including the starting BDF1 step.
A linear solve with MUMPS consists of an analysis (symbolic factorization), a factorization, and a solve step.
Because the mesh, and thus the nonzero pattern of the matrix, does not change throughout the simulation,
the symbolic factorization can be computed only once, and reused for the rest of the run,
so that each subsequent solve consists only of the factorization and solve steps.
The analysis step computes a row ordering limiting the fill-in, and can be computed
sequentially or in parallel. As it is computed only once, the time spent in the analysis step has little influence
on the overall computation time, however, the ordering depends on the chosen sequential or parallel strategy,
and affects the factorization times of all the subsequent solves.
In our tests, we found that the optimal factorization time for the range of problem sizes and number of cores used is achieved with a
parallel analysis using ParMetis \cite{karypis1997parmetis} in 2D, and with a sequential analysis using Scotch \cite{pellegrini2012scotch} in 3D.

The \emph{Fir} cluster on which the computations were run consists of nodes of two AMD EPYC 9655 CPU each, for a total of 192 cores per node.
Each CPU consists of 4 NUMA nodes with 3 Core Complex Dies (CCD) chiplets each, each with 8 cores.
To take advantage of these $2 \times 4 \times 3 = 24$ pools of cores sharing the same L3 cache,
the recommended configuration is a hybrid of 24 MPI ranks per node and 8 threads per rank\footnote{\url{https://docs.alliancecan.ca/wiki/Fir}}.
Our solver features distributed memory only (the assembly routines are not multithreaded),
but MUMPS supports hybrid MPI-OpenMP parallelism,
with multithreading used both by its own algorithms, and by the underlying BLAS library, which on \emph{Fir}
is AMD's AOCL, an implementation of BLIS optimized for AMD processors.
Since the solve time is expected to dominate the assembly time, especially in 3D,
a proper hybrid parallel setting can thus prove more efficient than a full MPI configuration,
as illustrated further.

\begin{figure}[hbtp]
  \centering
  \includegraphics[width=0.9\linewidth]{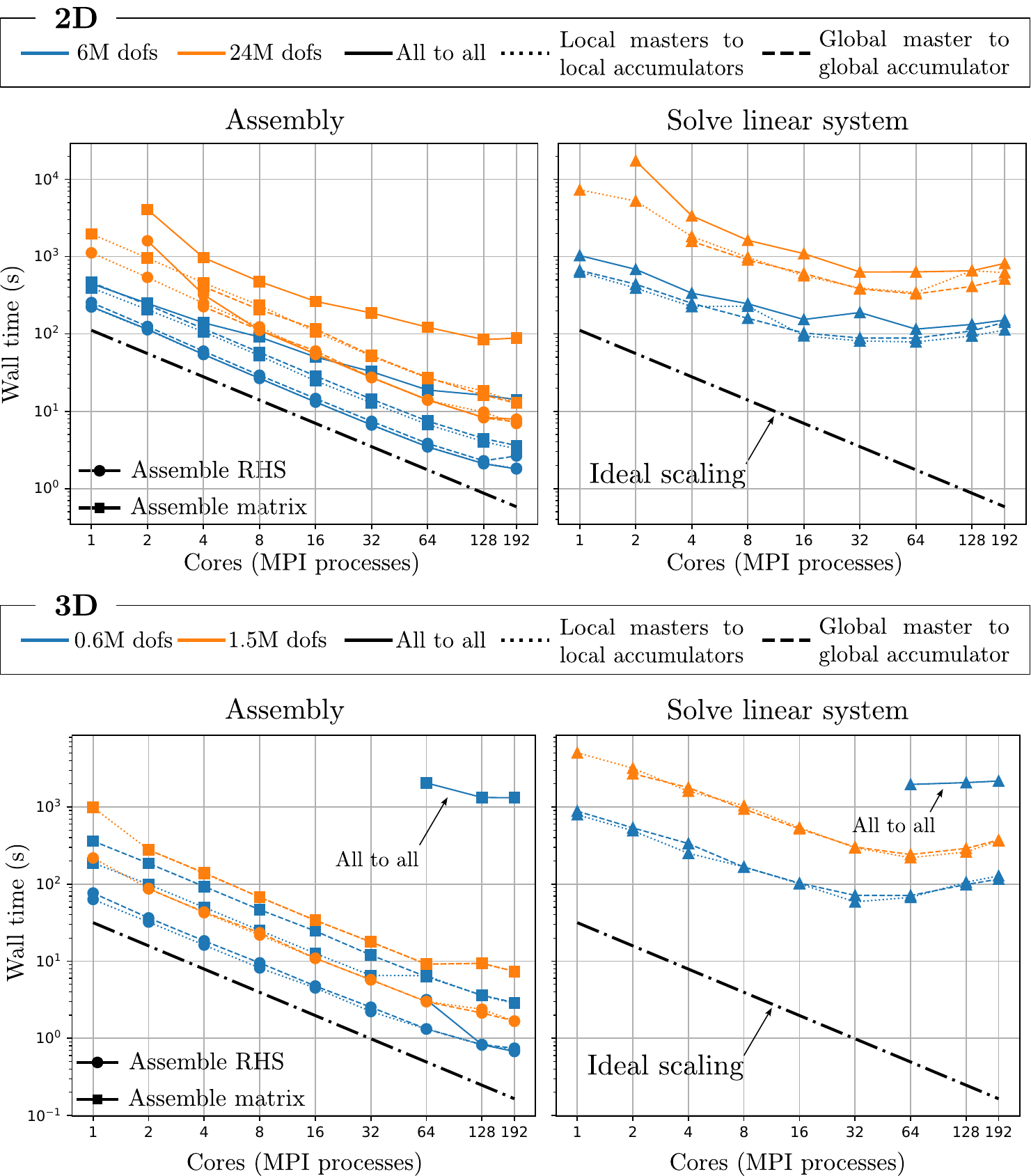}
  \caption{Comparison of the force-position coupling strategies in a full MPI configuration. Total assembly and solve times for 6 time steps for the 2D (top) and 3D (bottom) tests.}
  \label{fig:comparison_couplings}
\end{figure}

We start by comparing the three coupling strategies in a full MPI setting,
\Cref{fig:comparison_couplings}, for both 2D meshes, and for the coarse and medium 3D meshes only.
In all cases, the RHS assembly time is independent of the coupling strategy.
This is expected, as the chosen coupling strategy barely affects the RHS assembly.
The all-to-all strategy performs extremely poorly (plain curves), as expected as well:
compared to the local-to-local and global-to-global strategies,
the time assembling the matrix is about one order of magnitude greater in 2D, and 2.5 orders of magnitude in 3D on the coarse mesh.
The important overhead is spent applying the numerous force-position couplings to the PETSc distributed sparse matrix,
which involves a large amount of communication.
Similarly, the solve time is also directly affected by the additional couplings, especially in 3D.
All-to-all computations were not run on the medium mesh, as they would have been prohibitively long.
For the tested configurations, the local-to-local (dotted) and global-to-global (dashed) strategies behave
very similarly in terms of both assembly and solve times,
and are necessary to maintain acceptable computing times.
In the following we use the global-to-global scheme.

\begin{figure}
  \centering
  \includegraphics[width=0.88\linewidth]{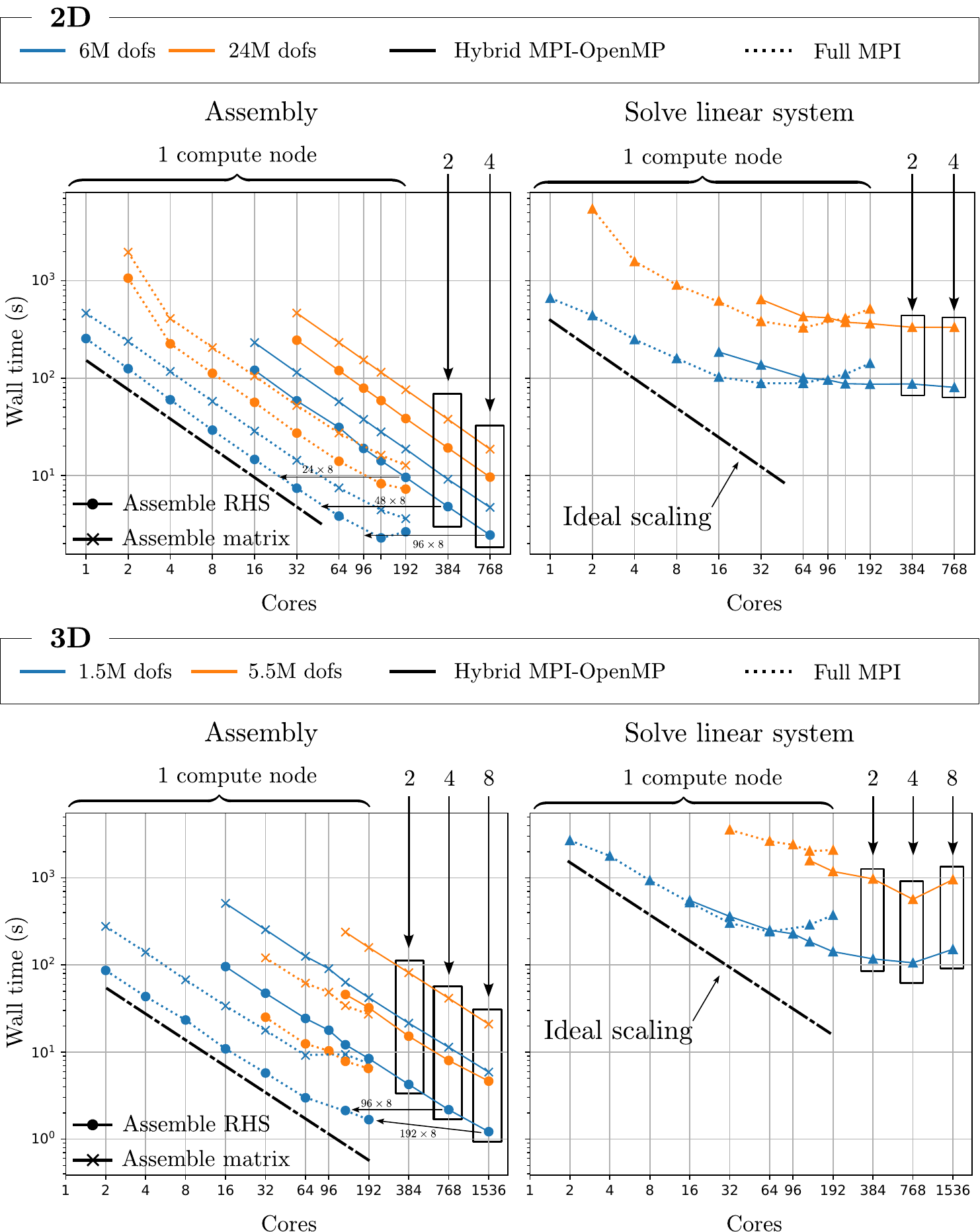}
  \caption{Scalings with full MPI and hybrid MPI-OpenMP configurations. Total assembly and solve times for 6 time steps on 1 to 8 compute nodes.}
  \label{fig:scalings}
\end{figure}

Lastly, we compare the scalings obtained with a full MPI and with a hybrid MPI/OpenMP configuration
for the FSI solver, using the global-to-global force-position coupling, \Cref{fig:scalings}.
The hybrid runs are controlled by the number of MPI ranks, and by the value of both
\texttt{OMP\_NUM\_THREADS} and \texttt{BLIS\_NUM\_THREADS}, which respectively set the number of threads
in the OpenMP regions and in AOCL.
In our tests, the default mappings and bindings chosen by \texttt{mpirun} proved
to be more efficient than any specified binding (e.g., by core, socket or L3 cache).
An example job script to submit through Slurm is provided in \ref{app1:job}.

The assembly routines are not multithreaded, and thus are expected to scale
with the number of MPI ranks only.
For all tests, the assembly time (including enforcing the FSI couplings) scales almost linearly with the number of MPI ranks,
that is, for up to 192 ranks in full MPI, and up to 96 ranks (4 nodes $\times$ 24 ranks)
in 2D (resp. 192 (8 nodes $\times$ 24 ranks) in 3D) in hybrid configuration.
In full MPI configuration,
the solve time only scales well up to a single node for the largest 3D test,
whereas for the smaller tests the optimal number of ranks is only 64,
after which the solve time plateaus or increases,
due to too little work assigned to each rank.
In hybrid configuration,
some scalability is achieved for up to 4 nodes for all tests,
ranging from a marginal gain for the 2D problems, to a reduction of the solve time
of about 1.7x for the fine 3D test (5.5M dofs, thus 1.375M dofs per node).

The linear solve remains the dominant limitation, especially
in 3D, where sparse matrix factorization produces much larger memory and
communication costs than in 2D.
These observations show that although a direct solver is suitable for prototyping,
a scalable iterative solver
and preconditioner are required to obtain larger 3D VIV simulations in an acceptable timeframe.

\section{Conclusions and future improvements}

We presented an implicit monolithic finite element method for the numerical
simulation of vortex-induced vibrations of two- and three-dimensional rigid cylinders
in the zero mass and damping limit.
In this formulation, the fluid forces are
recovered from a Lagrange multiplier enforcing the no-slip condition,
and are coupled to the mesh position on the cylinder to satisfy the
degenerate equation of motion of a massless solid body.
The fluid-structure interaction solver was verified
to assess the accuracy and consistency of the moving mesh ALE formulation, weak enforcement of boundary conditions,
and strong force-position coupling.
Second-order convergence was observed for the flow variables, fluid forces,
and mesh position, consistently with a BDF2 time-stepping scheme.
Then, fully distributed simulations of the VIV of rigid massless cylinders were presented.
Previously reported two-dimensional results obtained via the method of reactions
were reproduced, and extended to three-dimensional configurations, enabling simulations at Reynolds numbers of up to 300.
We obtain identical peak amplitudes of two-dimensional VIV oscillations in the periodic regime
compared to previous studies.

Three fluid-structure coupling strategies
were discussed to yield tractable assembly and solve times in a distributed-memory parallel environment.
For a rigid solid body, using position masters and force accumulators is necessary to limit the couplings between degrees of freedom.
Using a global-to-global coupling and MUMPS as linear solver, the proposed FSI solver was
shown to scale to up to 768 cores in a hybrid MPI/OpenMP configuration (96 $\times$ 8)
on 3D test cases of up to 5.5M degrees of freedom.
Relying on a direct solver is the current major bottleneck for larger-scale problems,
as it only allows dealing with up to a few million degrees of freedom.
To run simulations on finer meshes and at higher Reynolds numbers,
the next developments will focus on efficient preconditioning strategies
to enable iterative solvers for the fully coupled system.

\section*{Data availability}
\noindent The source code of the FSI solver is publicly available on GitHub (\url{https://github.com/arthurbawin/fez}).
The simulation data can be made available upon request.

\section*{Acknowledgments}
\noindent This work was supported by the Natural Sciences and Engineering Research Council of Canada (NSERC) through Discovery Grant RGPIN-2025-05020.
We gratefully acknowledge the Digital Research Alliance of Canada for the compute time on the \emph{Fir} cluster.
AB would like to thank A. Garon, N. Moës and S. Prudhomme for useful discussions on Lagrange multipliers,
as well as the \dealii community for help and advices to develop the proposed solver.

\section*{Declaration on generative AI in the writing process}
\noindent During the preparation of this work, the authors used ChatGPT to generate postprocessing scripts, as well as improve readibility of the manuscript. After using this tool, the authors reviewed and edited the content as needed, and take full responsibility for the content of the publication. 

\appendix
\section{Example Slurm script for hybrid MPI-OpenMP runs}
\label{app1:job}

\begin{lstlisting}[
  style=bashstyle,
  caption={Script to run FSI simulation with Slurm.},
  label={lst:run-script}
]
#!/bin/bash

#SBATCH --account=def-user
#SBATCH --nodes=1
#SBATCH --ntasks-per-node=24
#SBATCH --cpus-per-task=8
#SBATCH --mem-per-cpu=4000M
#SBATCH --time=0-01:00

export OMP_NUM_THREADS=8
export BLIS_NUM_THREADS=8

# MUMPS analysis for the 3D tests
ICNTL28=1
ICNTL29=2 % Parallel option, unused if ICNTL28=1
ICNTL7=3

srun /path/to/fez/build/monolithic_fsi param.prm \
    -mat_mumps_icntl_4 2 \     % MUMPS verbosity
    -mat_mumps_icntl_14 200 \  % Workspace size increase
    -mat_mumps_icntl_28 $ICNTL28 \
    -mat_mumps_icntl_29 $ICNTL29 \
    -mat_mumps_icntl_7  $ICNTL7 \
\end{lstlisting}

\bibliographystyle{cas-model2-names}

\bibliography{references}

\end{document}